\def\year{2022}\relax
\documentclass[letterpaper]{article} 
\usepackage{aaai22}  
\usepackage{times}  
\usepackage{helvet}  
\usepackage{courier}  
\usepackage[hyphens]{url}  
\usepackage{graphicx} 
\usepackage{natbib}  
\usepackage{caption} 
\DeclareCaptionStyle{ruled}{labelfont=normalfont,labelsep=colon,strut=off} 
\usepackage{algorithm}
\usepackage{algorithmic}
\usepackage{graphicx}

\usepackage{newfloat}
\usepackage{listings}
\floatstyle{ruled}
\newfloat{listing}{tb}{lst}{}
\floatname{listing}{Listing}

\usepackage{xcolor}
\newcommand{\answerYes}[1]{\textcolor{blue}{#1}} 
 
\newcommand{\answerNA}[1]{\textcolor{gray}{#1}}

\usepackage{booktabs}
\usepackage{enumitem}
\usepackage{subfigure}
\usepackage{soul}

\usepackage[most]{tcolorbox}
\usepackage{fontawesome5}

\newcommand\revision[1]{\bgroup\hskip0pt #1\egroup}

\newcommand{\dataset}[0]{Minerva-OSN} 

\newtcolorbox{cooltextbox}[1][]{%
    colback=black!5,
    colframe=black!5,
    notitle,
    sharp corners,
    borderline west={0pt}{0pt}{blue!80!black},
    enhanced,
    breakable,
    left=0pt,
    right=0pt,
    top=0pt,
    bottom=0pt
    }

\newtcolorbox{redtextbox}[1][]{%
    colback=red!10,
    colframe=red!15,
    notitle,
    rounded corners,
    enhanced,
    breakable,
    left=0pt,
    right=0pt,
    top=0pt,
    bottom=0pt
    }

\newtcolorbox{greentextbox}[1][]{%
    colback=green!10,
    colframe=green!15,
    notitle,
    rounded corners,
    enhanced,
    breakable,
    left=0pt,
    right=0pt,
    top=0pt,
    bottom=0pt
    }

\newcommand{\textbox}[1]{
    \noindent\fbox{%
        \parbox{0.97\columnwidth}{%
            {#1}
        }%
    }
}

\usepackage[most]{tcolorbox}
\newtcolorbox{LLMsnippet}{enhanced jigsaw,breakable, sharp corners,  colback=white,colframe=black, size=title, boxrule=0.5pt}

\title{Elephant in the Room: Dissecting and Reflecting on the Evolution of Online Social Network Research}

\author {
    Luca Pajola,\textsuperscript{\rm 1}
    Saskia Laura Schröer,\textsuperscript{\rm 2}
    Pier Paolo Tricomi,\textsuperscript{\rm 1}
    Mauro Conti,\textsuperscript{\rm 1}
    Giovanni Apruzzese\textsuperscript{\rm 2}
}
\affiliations {
    \textsuperscript{\rm 1} University of Padua, Padua, Italy\\
    \textsuperscript{\rm 2} University of Liechtenstein, Vaduz, Liechtenstein\\
    \{luca.pajola, mauro.conti\}@unipd.it, \{saskia.schroeer, giovanni.apruzzese\}@uni.li, pierpaolo@manalisk.it 
}

\begin{document}

\pagestyle{plain}

\maketitle


\begin{abstract}

Billions of individuals engage with Online Social Networks (OSN) daily. The owners of  OSN try to meet the demands of their end-users while complying with business necessities. Such necessities may, however, lead to the adoption of restrictive data access policies that hinder research activities from  ``external'' scientists---who may, in turn, resort to other means (e.g., rely on static datasets) for their studies. Given the abundance of literature on OSN, we---as academics---should take a step back and reflect on what we have done so far, after having written thousands of papers on OSN.

This is the first paper that provides a holistic outlook to the entire body of research that focused on OSN---since the seminal work by~\citet{acquisti2006imagined}. First, we search through over 1 million peer-reviewed publications, and derive \revision{13,842} papers that focus on OSN: we organize the metadata of these works in the \dataset\ dataset, the first of its kind---which we publicly release. Next, by analyzing \dataset, we provide factual evidence elucidating trends and aspects that deserve to be brought to light---such as the predominant focus on Twitter or the difficulty in obtaining OSN data. Finally, as a constructive step to guide future research, we carry out an expert survey (n=50) with established scientists in this field, and coalesce suggestions to improve the status quo---such as an increased involvement of OSN owners. \revision{Our findings should inspire a reflection to ``rescue'' research on OSN. Doing so would improve the overall OSN ecosystem, benefiting both their owners and end-users---and, hence, our society.}


\end{abstract}

\section{Introduction}

Online Social Networks (OSN) are platforms where \textbf{end-users} create virtual connections, share their experiences, communicate, and build social relations. Since the rollout of major OSN in the late 2000s -- such as Facebook and Twitter -- OSN have established a non-negligible space in the routines of billions of human beings~\citep{kim2018social}.
Indeed, we now use OSN \textit{also} for purposes revolving around aspects of our professional life. Examples include following and publishing news, finding work opportunities, purchasing or selling items, or even for full-time and OSN-specific jobs like ``social media manager'' and ``social influencer''~\citep{jin2019instafamous}.



The growth of OSN was simply impossible to overlook, and the \textbf{research community} did not stand idly by. Starting from the seminal work by~\citet{acquisti2006imagined} on Facebook, over the years thousands of scientific papers have been published, focusing on a plethora of themes revolving around various OSN and having plenty of societal implications. For instance, ~\citet{chu2010tweeting} started to design algorithms to detect ``bots'' in OSN, and years later,~\citet{bessi2016social} found that such bots played a crucial role in the outcome of the US elections. Since then, abundant papers have focused on detecting fake news~\citep{shu2017fake}. 

\par
Simply put, OSN are a (perpetually increasing) data source with infinite potential---and therefore are a valuable business for OSN \textbf{owners}. To sustain any OSN there is the effort of thousands of employees (at various levels), who work to provide a \textit{fulfilling experience to end-users} while \textit{sustainably generating revenue} for their own companies, and \textit{ensuring compliance} with modern norms (e.g., regulations and moral customs). Although some OSN owners may appreciate collaborating with researchers, some decisions may also put ``barriers'' that are difficult to overcome by academics---such as allowing access to OSN data only through thousands-of-USD-worth of fees~\citep{verge2023twitter}. 


The interplay among these three stakeholders (OSN owners, researchers, and end-users) delineates a complex ecosystem. Ideally, the aggregated contributions of every player should improve the ecosystem as a whole---which would ultimately lead to an overall enhancement of our quality of life. \textbf{As researchers, we want to facilitate such an improvement.} To this purpose, we ask ourselves a broad question: ``\textit{what has happened since the first research paper on OSN?}'' The OSN ecosystem has made enormous advances also thanks to researchers: in this context, are there some elements that have received a greater (potentially overshadowing others) degree of attention---and, if so, why? 



\begin{figure}[!ht]
     \centering
     \includegraphics[width=0.93\linewidth]{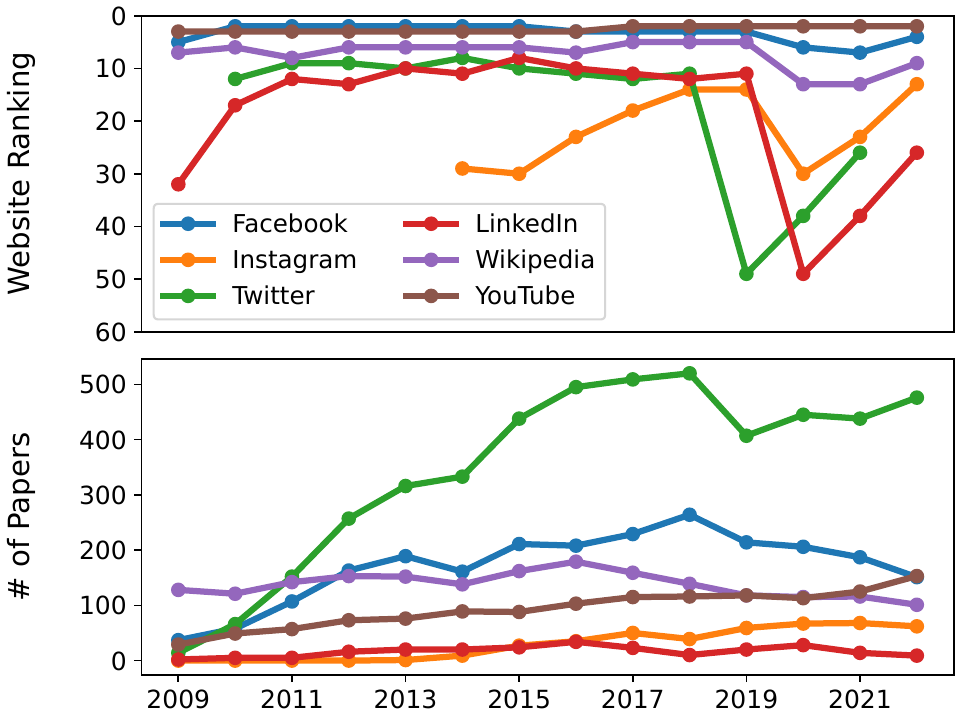}
     \caption{\textbf{OSN prevalence over time.} We highlight the discrepancy between OSN investigated in research and their real-world popularity (website ranking based on Alexa~\ref{link:alexa-top500}).}
     \label{fig:OSN-popularity-over-time}
 \end{figure}

In this paper, we perform a reflective exercise and seek to scrutinize \textbf{the entire body of OSN research}. Our goal is to examine whether prior literature, as a whole, conceals certain ``trends'' which, if not known, could be harmful to future research---and understand potential root causes. To this purpose, we first systematically collect the metadata of over 1 million publications, and identify \revision{13,842} works related to OSN. We organize these papers in the \dataset\ dataset, the first of its kind for OSN-research, which we publicly release. Next, through a meta-analysis, we investigate our \dataset\ dataset across four dimensions: \textit{OSN prevalence}, wherein we show that some OSN are significantly more studied than others \revision{(see Figure~\ref{fig:OSN-popularity-over-time})}; \textit{topic popularity}, wherein we reveal the topmost research topics considered in prior work; \textit{author analysis}, wherein we elucidate some aspects of the authors of the papers in \dataset; \textit{regulation and policies}, wherein we attempt to explain some of our findings in light of major changes, unrelated to research but which deeply affect OSN-related research.
Finally, to further validate some of the ``issues'' we uncovered, and to improve the status quo, we reach out to some of the authors of the papers in \dataset: through an (anonymous) expert survey, we collect the opinions of 50 renowned OSN researchers, highlighting the (often hidden) difficulties of carrying out impacting research in the OSN domain.

\paragraph{Contributions.}
We advance the state of OSN research by:
\begin{itemize}
\item Carrying out the largest meta-analysis of OSN-related literature, highlighting trends and aspects of prior research, and tying them with the OSN ecosystem. 
\item Releasing the dataset used for our meta-analysis, \dataset, comprising of \revision{13,842} papers (published between 2006 and 2023) included in 135 peer-reviewed venues and discussing \revision{91} OSN.

\item Conducting an expert survey with renowned OSN researchers (n=50), inquiring for their perspective on the present state of research and ways to enhance it.
\end{itemize}
All resources are publicly released in our repository.\footnote{https://github.com/pajola/Minerva-OSN} (Note: in the Appendix we provide a list of all non-bibliographic sources and their URLs---e.g., ~\ref{link:repository} refers to our repository.) 

\textbox{\textsc{\textbf{Positive Light}.} This paper is a reflective exercise in which we constructively scrutinize years of endeavors on OSN. Throughout our paper, we elucidate issues underpinning systematic problems---but which have not a single culprit. Hence, we are not pointing the finger at anyone.}

\section{Related Work}
\label{sec:related}

We are not aware of any paper whose contributions overlap with ours---despite thousands of works on OSN.

Most papers on OSN, such as~\citet{tang2022clues}, focus on a given phenomenon on one OSN; others, such as~\citet{fiesler2020no}, focus on a \textit{specific problem} across many OSN. Neither of these types of works attempts to provide a holistic view of OSN research.

Papers that analyze prior literature also tend to focus on \textit{one aspect}: for instance, the review by~\citet{andreassen2015online} considers works on OSN addiction, whereas the SoK by~\citet{singhal2023sok} focuses on content moderation. \revision{Similarly, various reviews considers a single OSN: this is the case of~\citet{antonakaki2021survey} for Twitter; and of~\citet{wilson2012review} for Facebook.}  In contrast, our analyses are agnostic to the focus of any analyzed paper (as long as it pertains to OSN) \revision{and we do not restrict our vision to any specific OSN.}

Furthermore, many papers collect the opinion of experts to guide future research---either through questionnaires, such as~\citet{dwivedi2023so}; or through interviews, such as~\citet{bieringer2022industrial}. However, we are unaware of any work that carried out an expert study in the OSN context.

Finally, to derive our conclusions, we analyze the metadata \revision{(e.g., titles and abstract)} of prior work. Such an approach is widely adopted in some research domains -- e.g., management and business~\citep{donald2022content, gaur2018systematic} -- but it has never been used before in the OSN domain. \revision{This is why our research also encompass original methods. 
For instance, we combine automated ``AI-driven'' techniques with manual ``human-driven'' analyses. As a practical example of our technical innovation, we: {\small \textit{(i)}}~use a topic model to determine which topic best describes any given paper, {\small \textit{(ii)}}~augment the model's accuracy via our domain expertise, and {\small \textit{(iii)}}~systematically validate the model's quality via a tri-party reviewer system---we are not aware of works adopting a similar quality assurance procedure.}

\section{The Minerva-OSN Dataset}
Among our major contributions, we introduce the \dataset~dataset. \revision{(Minerva was the goddess of wisdom in Roman mythology.) We use the dataset as a basis to derive our findings. Here, we first elucidate the methodology employed in the creation of \dataset{} dataset---which is rooted in the guidelines of systematic literature reviews~\citep{booth2021systematic}. Then, we explain how we carried out the analysis on topic modeling---representing a pivotal part of our research and a substantial technical contribution in this domain.
}

\vspace{-1mm}

\subsection{Data Collection (Creation of Minerva-OSN)}

We design \dataset\ with the intention of collecting ``all'' \textit{OSN-related} and \textit{peer-reviewed} papers (in computer science) that have ever been published. Hence, the first step entails identifying research-focused venues (i.e., journals and conference proceedings) that contain such publications.

\revision{\textbf{Venues and candidate papers.}}  We begin our search with Google Scholar (among the most well-known repositories of scientific articles), and consider the subcategories within the ``Engineering and Science'' category~\ref{google-scholar-cats}. For each subcategory, Google Scholar lists the 20 most popular venues.  
We integrate these 20 top-tier venues with all those listed in the Springer's Lecture Notes in Computer Science and the International Conference Proceedings Series of the Association of Computing Machinery. We manually review all identified venues, remove duplicates and merge venues that changed publisher or name (e.g., WWW changed to TheWebConf). In total, we identify 135 relevant venues \revision{(reported in the Appendix)}. 
Then, for each venue, we query the Scopus database for \textit{all} papers that appeared therein---starting from 2006 (the year in which Facebook was released to the general public)\revision{, as also done by~\citet{antonakaki2021survey}}. For every paper, we collect its metadata, i.e. title, authors (and affiliations), abstract, publication date, DOI. In total, we extract over 1 million ``candidate'' papers.

\textbf{Identifying papers on OSN.} There is no straightforward way to determine if any given paper is related to \revision{(any)} OSN. Hence, we design a sensible \revision{(and original)} \textit{heuristic}: for each paper (among our 1+ million), we automatically analyze its abstract, looking for \textit{mentions of the name of at least one OSN} (intuitively, a paper on OSN ``should'' mention an OSN in its abstract). \revision{Applying our heuristic requires us to determine a list of OSN that we can use to check the abstracts. To this end, we are inspired by the procedure followed by~\citet{fiesler2020no}, and rely on the Wikipedia page on OSN~\ref{link:wikipedia-osn-list}; moreover, to provide a more comprehensive coverage, we expand such a list by including alt-tech OSN~\ref{link:wikipedia-alttech-osn-list} as well as YouTube---which, surprisingly, was not considered by~\citet{fiesler2020no} despite its popularity. Overall, we identify 296 unique OSN that we use to apply our heuristic,\footnote{\revision{We also compiled a list of OSN synonyms; for instance, ``Sina Weibo'' is often referred to just as ``Weibo''~\citep{yang2022save}.}} leading to 15,770 papers.

\textbf{Filtering and Validation.} Our heuristic may yield some false positives. To provide more accurate results, after removing duplicates, we excluded all papers published before the rollout of the ``matching'' OSN (e.g. ``Discord'' was released in 2015, and any article mentioning ``discord'' in the abstract before 2015 could not be on this OSN). Then, we scrutinize ``homonyms,'' i.e. words that can be used to denote an OSN, or a completely different concept (e.g., ``band''): in these cases, we only include the paper if the abstract also mentions other common terms on OSN (e.g., ``social network''). Lastly, we perform a manual inspection: for each OSN (out of 296), we inspect up to 50 matched papers, discarding those that, despite meeting the above-mentioned criteria, do not focus on OSN. Such a procedure is useful to remove papers on less-known OSN (e.g., in one abstract, the OSN ``Ning'' was associated with a person named Ning): out of 74 excluded papers, 73 pertained unpopular OSN.} 


\textbox{\textsc{\textbf{Summary}.} The \dataset{} \revision{after filtering contains 13,842 OSN-related papers published in 2006--2023 across 135 venues.} 
Our \dataset\ is the largest curated dataset of this kind (and adheres to FAIR principles~\ref{fair}).}

\vspace{-1mm}

\textbf{Limitations.} We cannot claim that Minerva-OSN contains \textit{all} publications on OSN. Some papers may not be indexed in Scopus. Moreover, our selection of venues, despite its comprehensiveness, may have omitted some journals or conferences. Finally, our heuristic may have not captured papers that did not mention any specific OSN in the abstract.


\subsection{Topics Identification (Technical Analysis)}
\revision{To derive some of our findings, we leverage simple scripts (e.g., cross referencing among various papers) applied to \dataset{}. However, one of our goals cannot be feasibly achieved via human effort alone: to identify the topmost ``topics'' of the 13,842 papers in \dataset{}, we rely on artificial-intelligence techniques---and, particularly, on \textit{topic modeling}~\citep{dieng2020topic}. Yet, contrarily to most prior work -- also in the OSN context, e.g.,~\citet{sartori2023impact} -- we do not ``blindly'' apply such AI-based techniques: we also {\small \textit{(i)}}~guide the topic model with our own expertise in the OSN domain, and {\small \textit{(ii)}}~conduct a thorough manual analysis focused on assessing the results of our topic model. Let us explain our procedure, which is systematic in nature. 


\textbf{Topic modeling.}
The main idea of topic modeling is to use a model to identify ``topics'' (i.e., thematic patterns) in a collection of documents---which, in our case, are the abstracts of the papers in \dataset{}.
Our model of choice is the well-known BERTopic~\citep{grootendorst2022bertopic}, which we apply through the following steps: 
{\small \textit{(i)}}~we create a sentence embedding for each abstract by using ``bge-small-en-v1.5'' (one of the best open-source models~\ref{link:Huggingface leaderboard});
{\small \textit{(ii)}}~apply dimensionality reduction to such sentence embeddings via UMAP~\citep{mcinnes2018umap};
{\small \textit{(iii)}}~cluster the reduced sentence embeddings via HDBSCAN~\citep{campello2013density};
{\small \textit{(iv)}}~create topic representations based on the resulting clusters, i.e., we identify the \textit{most relevant words describing the topics clustered}, which we do by using Count Vectorizer, removing English stopwords, and applying c-TF-IDF; {\small \textit{(v)}}~finally, we assign a label on the resulting topics by using LLAMA-2~\citep{touvron2023llama}.\footnote{\revision{\textbf{Trial-and-Error and Alternatives.} The procedure described above was derived after extensive testing of various alternatives. E.g., for step {\scriptsize \textit{(i)}}~we considered also ``all-mpnet-base-v2'' and ``bge-small-en''; 
for {\scriptsize \textit{(ii)}}~we considered also PCA; for {\scriptsize \textit{(iii)}}~we considered also K-Means; and for {\scriptsize \textit{(v)}}~we considered KeyBert and MMR.}}}
\revision{
Indeed, as recommended by~\citet{grootendorst2022bertopic}, we use Large Language Models (LLM) to facilitate the human-interpretability of our findings. Specifically, we use LLM in two ways. First, as we mentioned, we use LLAMA-2 for \textit{topic labeling} (i.e., ``given a set of keywords that denote a topic, what is a name that can denote them?''); we do this by issuing the prompt1 reported in the Appendix. 
Second, we use GPT-4 to \textit{describe the content} of each topic identified by BERTopic (i.e., ``given the keywords and the label of a topic, what is a human-readable description of such a topic?''); we do this by issuing the prompt 2 reported in Appendix. 
}

\begin{figure*}
    \centering
    \includegraphics[width=0.95\linewidth]{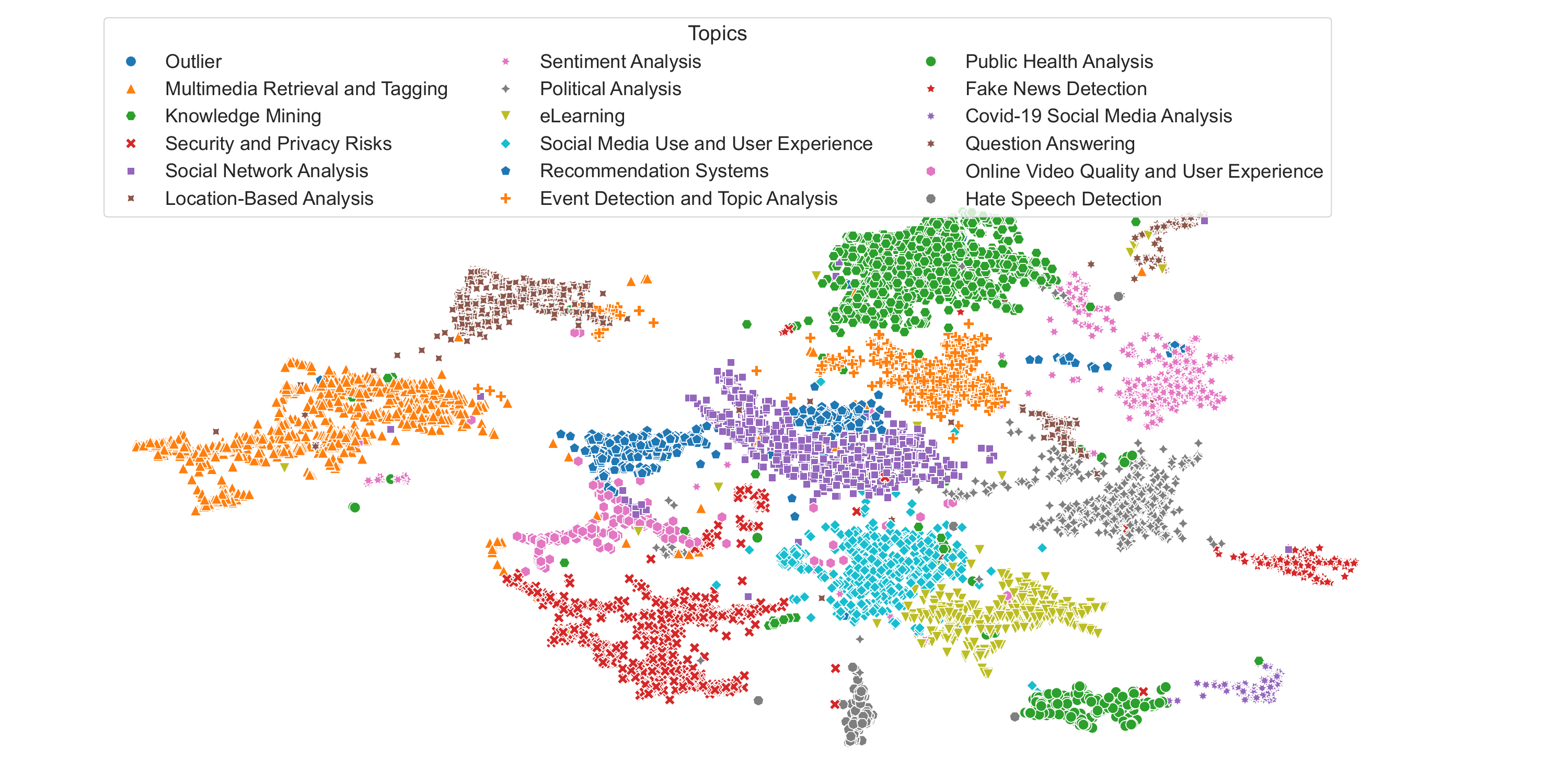}
    \caption{BERTopic visualization.}
    \label{fig:bertviz}
\end{figure*}

\revision{\textbf{Human guidance and Results.}
We found that the ``blind'' application of the abovementioned workflow led to misleading results.\footnote{\revision{For instance, the topic model was assigning hundreds of papers to a single cluster labeled ``Twitter topics'', which was not useful.}} Hence, we use our domain expertise to guide the topic modeling in providing informative results. Specifically,
to prevent the topic model from being biased towards any OSN (typically, popular ones) and to cluster abstracts only based on their actual contents, we replace the specific OSN name of each abstract with a neutral tag (e.g., we replace ``Twitter'' with ``OSN''). After applying such a preprocessing step and re-doing the operations discussed in the previous paragraph, our topic model assigned a topic to 83.9\% papers in \dataset{}. Specifically, 11,606 papers were associated to one among 17 topics (automatically identified by our model). Such an outcome is in line with observations of prior work~\citep{karabacak2023natural}, since the presence of ``outliers'' is expected in such a sparse collection of documents.\footnote{\revision{We will still consider these 2,236 outliers in the remainder of our paper---albeit not for topic-related analyses (given that our focus is on the ``topmost'' topics studied by prior OSN literature).}}
We report in Figure~\ref{fig:topicList} distribution of the papers in \dataset{} across the 17 topics identified by our topic model; in the Appendix, we report the description of each topic (Table~\ref{tab:topic}) and a visualization of our clustering (Figure~\ref{fig:bertviz}). However, before we present these findings, we must ascertain the quality of our topic model: this is necessary to ensure that our following analysis is well founded.

\textbf{Tri-party validation.}
There is no ground truth available that we can use to ``automatically'' measure the accuracy of our topic model---e.g., we do not know (a priori) if a paper put in the ``eLearning'' cluster is indeed about eLearning. Hence, following the recommendations by~\citet{hagen2018content}, we must inspect the results ourselves. However, we found no prior work that systematically assessed the combination of BERTopic and LLAMA2 for the specific purpose of topic modeling for abstracts of OSN-related papers. Actually, most related work (e.g.,~\citet{koloski2024aham}) merely mention some manual review process but without describing the exact methodology. Therefore, to provide a foundation for future work, we designed an original quality-assurance pipeline that encompasses a tri-party (human-driven) reviewing system. 
Specifically, we proceed as follows.
First, for each of the 17 topics identified by our model, the abstracts of 10 different papers are randomly selected.
Next, these 170 abstracts are provided to three experts with an academic background (in OSN and AI): these experts \textit{do not know} which topic was assigned to each abstract by the topic model, but they know the list of possible topics (in Table~\ref{tab:topic}) and their description. 
The experts must then independently select two topics that they think better capture the content of each abstract, specifying their first and second choices. 
Afterward, we assess {\small \textit{(a)}}~how much the experts agreed among each other, and {\small \textit{(b)}}~how much the experts agreed with the topic model. The results are positive. Specifically, for {\small \textit{(a)}}, we found a remarkable 73.7\% agreement among the first choice of each reviewer---denoting that the reviewers share similar views~\citep{o2020intercoder}. For {\small \textit{(b)}}, we found that the topic assigned by BERTopic was placed first by the reviewers in 72.5\% of the cases, and first or second in the 80.2\% of the cases. We further examined the responses, scrutinizing major occurrences of agreement/disagreement. We found that for more specific topics like ``Covid-19 Social Media Analysis'' and ``Hate Speech Detection'' the consensus was higher (first choice in~$>$90\% of the cases). Conversely, broader topics such as ``Multimedia Retrieval and Tagging'' or ``Online Video Quality and User Experience'' lead to greater disagreement (accuracy slightly~$>$50\%). However, even in these challenging cases, we appreciate that the agreement increases by considering also the second choice (accuracy $>$70\%). We can hence conclude that the topics assigned by our topic model are, in general, appropriate---thereby validating our future analyses.

\begin{figure}[!t]
    \centering
    \includegraphics[width=0.95\linewidth]{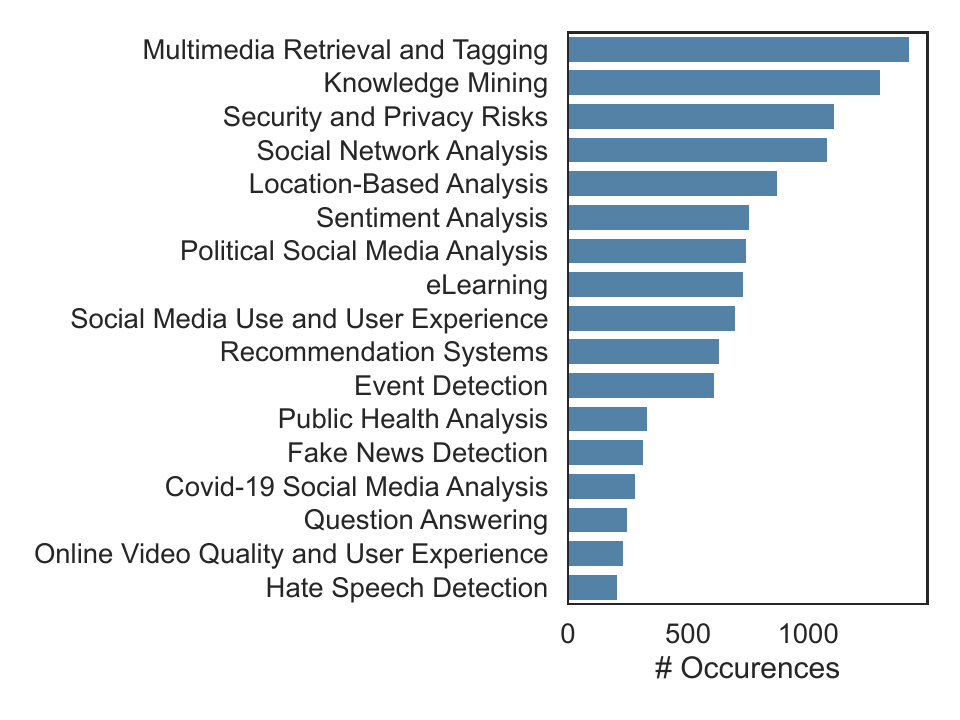}
    \caption{\revision{\textbf{Distribution of topics in \dataset{}.} After processing the 13,842 papers in dataset, our topic model identified 17 topics which described the abstracts of 83.9\% of the papers in \dataset{} (topic description in Table~\ref{tab:topic}).}} 
    \label{fig:topicList}
\end{figure}

\textbox{\textsc{\textbf{Findings} (topic prevalence).} 
From Figure~\ref{fig:topicList}, we see \dataset{} papers span across a variety of topics. 
The three most prevalent topics are ``Multimedia Retrieval and Tagging'' (1,426 papers), ``Knowledge Mining'' (1,305), ``Security and Privacy Risks'' (1,110). This result is somewhat expected (e.g., also~\citet{antonakaki2021survey} mention many of the themes captured in Figure~\ref{fig:topicList}) but we are the first to quantify such prevalence in the whole body of OSN literature. 
}

\vspace{-1mm}


}

\section{Analyses (``What Have We Done So Far?'')}
We now focus on our \revision{primary} contribution and analyze the \dataset\ dataset to derive \revision{novel findings on OSN research. We begin by investigating \textit{generic research trends}~(\texttt{RQ1}), and then focus on the \textit{authors} of prior works~(\texttt{RQ2}). Next, via deeper analyses, we elucidate the diversity of topics studied in specific OSN~(\texttt{RQ3}). Then, inspired by the advent of country-specific regulation such as the GDPR, we will examine potential differences from a geographical perspective (\texttt{RQ4}). Finally, we explore the relationship between an OSN real-world popularity and the degree of research activity devoted to such an OSN~(\texttt{RQ5}).}


\begin{figure}[!t]
    \vspace{-5mm}
    \centering
    \includegraphics[width=.85\linewidth]{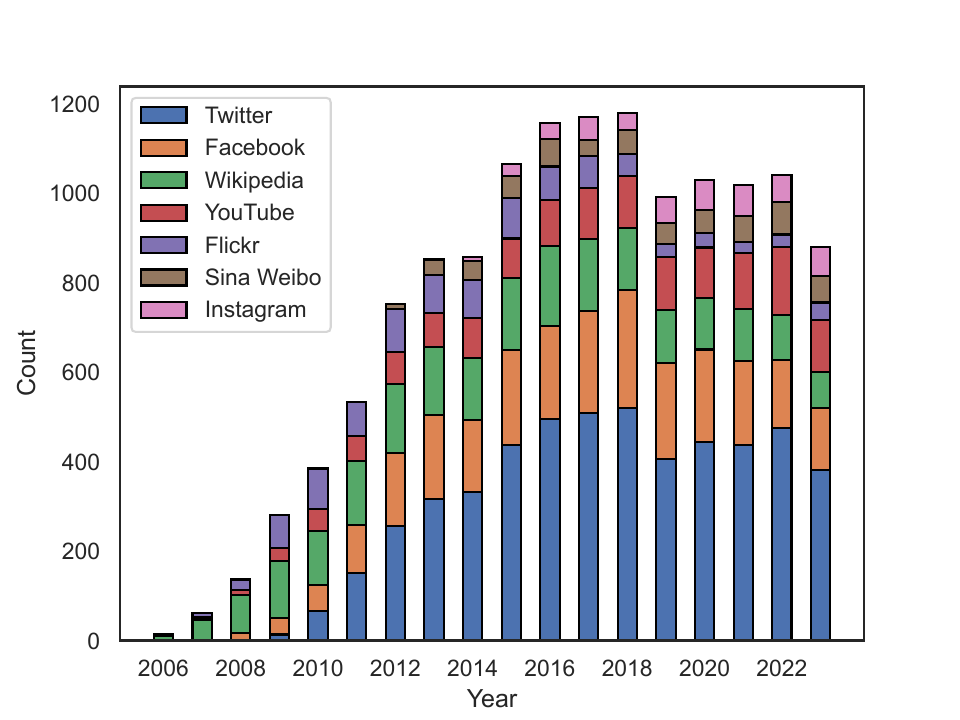}
    \caption{\revision{Yearly distribution of the most prevalent OSN.}}
    \label{fig:osn-dist}
\end{figure}

\subsection{\texttt{RQ1}: Major OSN Research Trends}

\textit{What has OSN literature mostly focused on in 2006--2023?} 
\revision{We extend our preliminary analysis (focused on the topmost topics) on \dataset{} to gain a better understanding of major research trends over time. Such a temporal analysis represents the cornerstone of our paper. To answer this RQ, we define the two following sub-RQs:}
\begin{enumerate}[label=\texttt{RQ1.\alph*:}, leftmargin=2em, align=left, noitemsep]
    \item \textit{Temporal OSN prevalence}, where we analyze the most researched OSN over time (2006--2023). 
    \item \revision{\textit{Topics over Time}, where we investigate how the main topics evolved throughout the years 2006--2023.}
\end{enumerate}



\paragraph{Methodology.}
For \texttt{RQ1.a}, we analyze the abstracts of each paper in our dataset and identify the OSN considered. We aggregate the results on a yearly basis and create a stacked histogram plot. 
\revision{For \texttt{RQ1.b} we investigate how the 17 topics found by our topic model have evolved over time by analyzing the yearly frequency of their occurrences.}
\par

\paragraph{Results.}
[\texttt{RQ1.a}]
Overall, of the \revision{296} OSN we considered when creating \dataset, only \revision{91 (30.6\%)} have been investigated by academic work. Among these, 77 OSN have been researched by less than 100 papers, and \revision{52} OSN have even been studied by less than ten papers (we report these details in \revision{the Appendix}). 
Figure~\ref{fig:osn-dist} shows the evolution of the seven most prevalent OSN. In general, the peak of OSN research activity was in 2018 and has been slightly decreasing since.\footnote{The observed decline in the number of research publications in the year 2023 is due to Scopus not being updated with the latest publications of 2023 (we created \dataset\ in Dec. 2023).} 
Twitter is the most prevalent OSN in research, followed by \revision{Facebook, Wikipedia, and YouTube}. We also observe that Twitter has been predominant in OSN research since 2012 and that Flickr's share has decreased since 2018.\footnote{This might be related to the fact that Flickr announced the end of its 1 TB free storage plan in November 2018~\ref{link:flickr-1TB}.} This clearly highlights that \textit{OSN research is not equally distributed across different OSN but concentrates on a few}, with Twitter being predominant. \revision{Even though it was known that Twitter was well-studied in research~\citep{antonakaki2021survey}, we are the first to provide quantitative results on the whole OSN panorama.}



\begin{figure}[!h]
     \centering
     \includegraphics[width=.95\columnwidth, keepaspectratio]{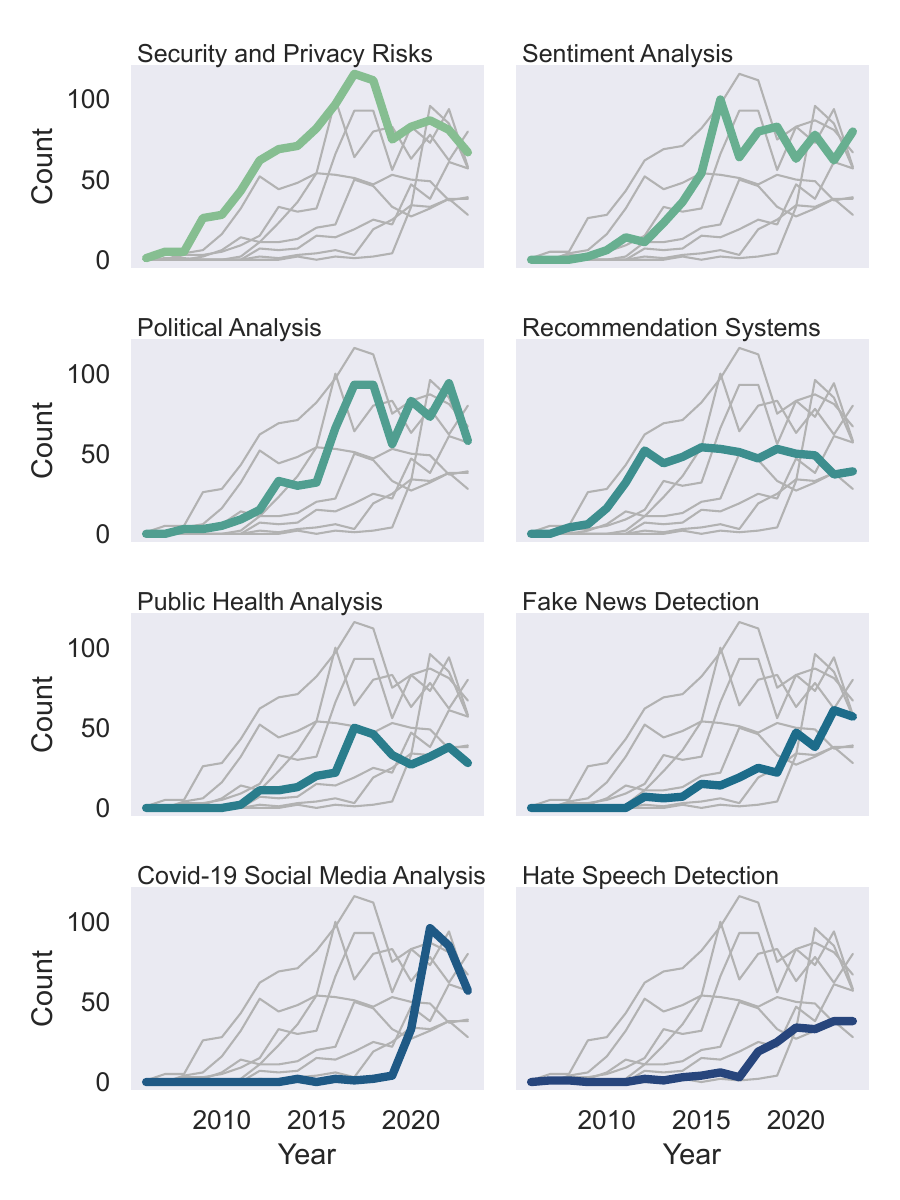}
        \caption{\revision{\textbf{Topics over Time} (for 8 topics). The y-axis reports the number of papers on each topic. Thick lines show the trend of that specific topic, thin lines show all the others.}}
     \label{fig:topics-over-time}
 \end{figure}

%
%
[\texttt{RQ1.b}] Figure~\ref{fig:topics-over-time} reports the evolution of topics across OSN over time. 
\revision{Different trends can be observed for each specific topic. 
For instance, consider ``Covid-19 Social Media Analysis'': this topic emerged in 2020,} peaked in 2021, and has declined since (this expected result confirms the validity of our prior analyses).
\revision{There are also emerging topics that have found continuous interest from the community. 
This is the case of ``Hate Speech Detection'' and ``Fake News Detection'' for instance.
Conversely, topics like ``Sentiment Analysis'' and ``Security and Privacy Risks'' are topics that have been always studied by the community. 
These findings suggest a community that adapts its interest frequently with the evolution of different trends of OSN mass utilization. 
}

\begin{greentextbox}
{\scriptsize \faChartLine } \textbf{\texttt{RQ1}: What has OSN literature mostly focused on in 2006--2023?} 
There is a tendency to focus on a few OSN, Twitter being the most popular. Researchers study OSN for different reasons, ranging from sentiment analysis and recommendation systems to sensitive topics such as hate speech detection. Interest in these topics follows temporal trends that vary---likely tied to real-world attention. 
\end{greentextbox}

\begin{figure}[t]
    \vspace{-4mm}
    \centering
    \includegraphics[width=.85\linewidth]{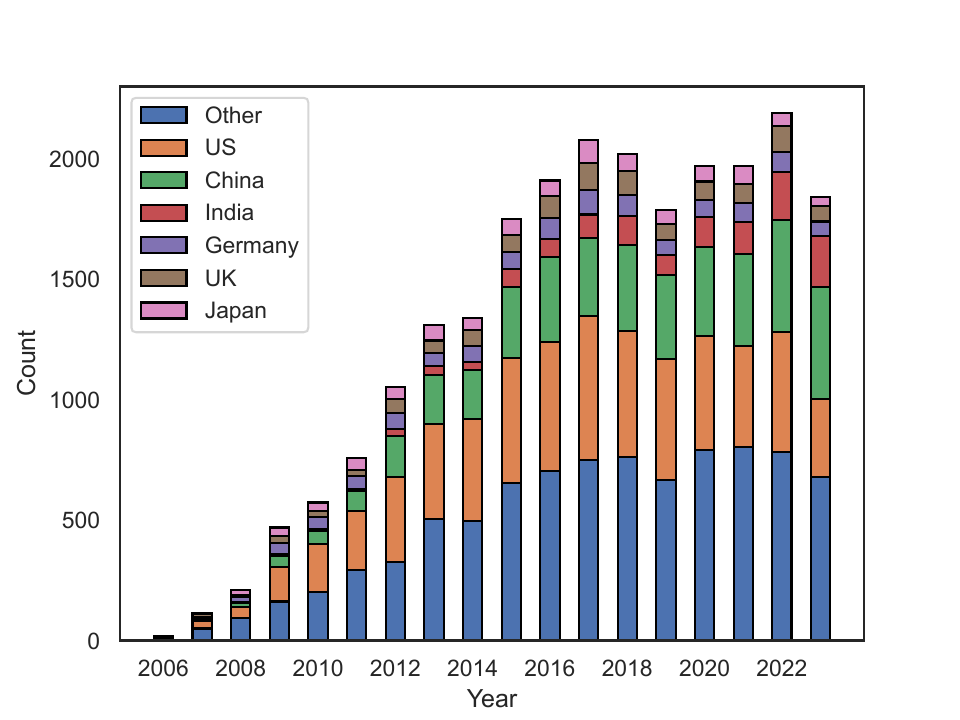}
    \caption{\revision{Authors' affiliation country (yearly distribution).}}
    \label{fig:countries}
\end{figure}

\subsection{\texttt{RQ2}: Analysis of the Authors of OSN Papers}
\textit{What are the predominant patterns among authors researching OSN?}
With this research question, we investigate the traits of scientists contributing to the field of OSN. In this way, we provide another perspective on the state of research. 
We disentangle this RQ with the following analyses:
\begin{enumerate}[label=\texttt{RQ2.\alph*:}, leftmargin=2em, align=left]
    \item How has the geographical distribution of authors evolved over time?
    \item Do OSN researchers tend to interact directly with the platforms that are studied in their papers?
    \item How diverse are the topics studied by OSN researchers?
\end{enumerate}
Through this RQ, \revision{we seek to examine if OSN research is becoming more inclusive or exclusive} to specific communities. 


\paragraph{Methodology.}
We answer \texttt{RQ2} by analyzing the author's information contained in our dataset. We recall that Scopus provides some author information for each indexed paper. 
To answer \texttt{RQ2.a}, we retrieve general statistics such as the total number of authors per year and the country of their affiliation. 
For \texttt{RQ2.b}, we use regular expressions to identify whether any co-author of a paper is working for a specific OSN.\footnote{A limitation of this approach is that an author may be affiliated to a given OSN, but the affiliation reported on the paper is a different one. Hence, our numbers may underestimate the reality.} We report the percentage of papers focusing on an OSN with at least one author affiliated to such an OSN.
For \texttt{RQ2.c}, we conduct a cross-analysis of the authors w.r.t. the 17 topics found by the topic model. We examine how many unique topics are (on average) studied by each author. 

\paragraph{Results.}

[\texttt{RQ2.a}] We analyze the contributions of each country to OSN research from 2006 to 2023 by examining the affiliations of the authors for each paper. Figure~\ref{fig:countries}, \revision{shows the 6 most represented countries (we aggregate countries outside the top-6 in an ``other'' bin; the detailed ranking is in the Appendix).
We observe that the US is the leading country in terms of quantitative contributions since the dawn of OSN research. In contrast, for China and India, the number of OSN papers has increased over time (especially recently). Other countries like the UK and Germany provide a stable contribution.} 
Additionally, we record the number of unique authors who have published at least one OSN paper per year according to \dataset{}. The results show a growing interest in the research community: while in 2006 only \revision{48} researchers focused on OSN, the number increased to \revision{2,707} in 2014; and reached the maximum of \revision{4,111 in 2022}. After 2016, the number of unique authors remains stable around \revision{3,900} per year.
This finding suggests that OSN research, after over 10 years of growth, reached a plateau after 2016. 


[\texttt{RQ2.b}] With respect to authors affiliated with OSN companies, we observe the following:
\revision{5.2\%} of papers focusing on Facebook or Instagram have at least one author affiliated with \revision{(i.e., employed by)} Facebook or Instagram (in total \revision{2,860} papers focus on Facebook and Instagram). Twitter affiliates are listed as authors in \revision{2.0\%} of the \revision{5,248} papers on Twitter. Wikimedia (Wikipedia) affiliates are authors in \revision{1.5\%} of \revision{2,139} papers. 
We also identify an interesting trend on LinkedIn:\footnote{We did not include Microsoft affiliations since LinkedIn was only acquired by Microsoft in 2016.} \textit{almost \revision{16.6}\% of 253 papers on LinkedIn have at least one co-author affiliated with LinkedIn}.
\revision{Nevertheless, we also found that the number of publications co-authored by an Instagram or Facebook employee is twice as high as those co-authored by a Twitter employee: this is intriguing given that the number of research papers on Twitter is twice as large as that of Facebook or Instagram (see Figure~\ref{fig:OSN-popularity-over-time}).
Such a phenomenon is probably driven by Twitter's API being (until 2023~\ref{twitter-api}) freely available to use, thereby enabling ``anyone'' to obtain Twitter data usable for research on Twitter.} 
\revision{Overall, these trends suggest that research on a specific OSN, such as LinkedIn, is only accessible with the direct involvement of the OSN---highlighting the relevance of ``data-access policies'' enforced by OSN owners (we will elaborate on this aspect in the next section).}
\par
[\texttt{RQ2.c}] Nearly \revision{32,000} distinct researchers have co-authored papers included in \dataset\ (notably, \revision{2.6}\% wrote more than five papers on OSN).
We analyze how many topics are studied by these authors (on average). We now report statistics for authors with varying impact, i.e., for authors that contributed to at least $i$ publications.
For $i=1$, the average is 1.2 topics; for $i=5$, it increases to \revision{3.6}; for $i=10$ and $i = 15$, the average number of topics is \revision{5.3} and \revision{6.5}, respectively. 
This finding indicates that authors predominantly focus their efforts on specific research domains, and that more \revision{experienced} authors tend to expand their coverage of topics eventually---which is a positive result.

\begin{greentextbox}
{\scriptsize \faChartLine } \textbf{\texttt{RQ2}. What are the predominant patterns among authors researching OSN?} 
The US leads OSN research. Some platforms (e.g., LinkedIn) have more papers co-authored by their employees. Research tends to focus on a few topics, but authors broaden their scope over time.
\end{greentextbox}



\subsection{\texttt{RQ3}: Topic Distribution Across OSN}
\textit{How do research topics differ across OSN?}
Motivated by our previous findings showing that most academic work focuses on a few OSN, we want to further understand the relationship between research topics and OSN. We examine whether certain topics are primarily addressed using a single OSN (indicating strong dependency) or investigated across multiple OSN (weak dependency). \texttt{RQ3} has no sub-RQs.

\begin{figure}[!t]
    \centering
    \includegraphics[width=\linewidth]{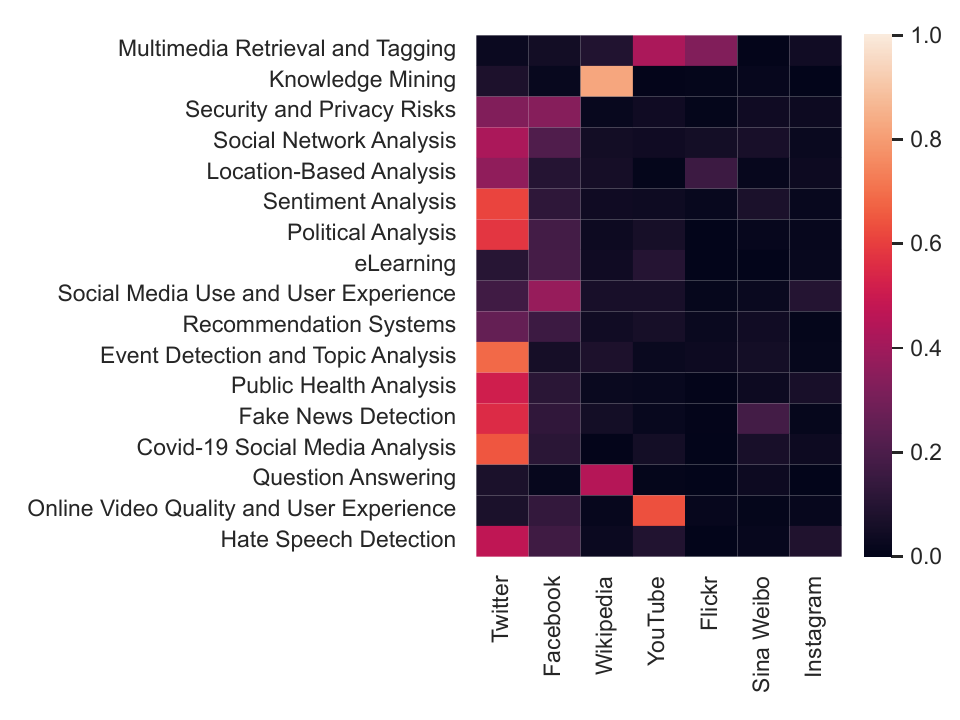}
    \caption{\revision{Topic distribution across 7 most prevalent OSN.}}
    \label{fig:osn_topics}
\end{figure}

\paragraph{Methodology.}
We conduct this analysis considering the 17 topics identified in \texttt{RQ1.b}. Recall that our topic model will yield one topic for each paper in our \dataset\ dataset; at the same time, each paper is associated to one or more OSN. We then create a function that maps every topic to all the OSN for which there is at least one paper that focuses on that topic. Next, we count the number of unique OSN associated to each topic. Finally, each topic distribution is normalized between 0 and 1. 
\paragraph{Results.}
Figure~\ref{fig:osn_topics} illustrates the distribution of topics across the seven OSN with the highest number of papers in \dataset\ (see Table~\ref{tab:current-osn-popularity}).
Intriguingly, Twitter, the most prevalent OSN, monopolizes research topics, reaffirming its foundational role in OSN research.
Facebook appears to be relevant on topics such as \revision{``Social Media Use and User Experience,'' and ``Security and Privacy Risks.''} Similarly, Wikipedia appears to be strongly related to topics such as \revision{``Knowledge Mining,'' and ``Question Answering.''} 
Also, we note that there are some topics that are not investigated by any of these seven OSN. For instance, \revision{``eLearning''} is strongly connected to Moodle. 
These results show that some research topics are strictly dependent on specific OSN.


\begin{greentextbox}
{\scriptsize \faChartLine } \textbf{\texttt{RQ3}. How do research topics differ across OSN?}
Most research topics are highly dependent on a specific OSN. \revision{Twitter is the OSN with the highest variety of topics; Wikipedia and YouTube dominate on two specific topics, while Flickr dominates on one topic---in common with YouTube (``Multimedia Retrieval and Tagging'').}
\end{greentextbox}

\subsection{\texttt{RQ4}: \revision{OSN Research Across the World}} 
\revision{\textit{How does the OSN-research output differ across geographical areas?}
This RQ has a twofold purpose. On the one hand, we want to investigate if there are some notable disparities in the OSN research pursued in different areas of the World. On the other hand, we want to use our analysis as a proxy to investigate the potential impact that the GDPR had on OSN research.\footnote{\revision{\textbf{Background.}
The GDPR (General Data Protection Regulation) 
is a European regulation released in April 2016, and implemented in May 2018 to protect natural persons in regard to the processing of their personal data. The GDPR inevitably also influenced OSN owners, requiring them to increase user protection and asking for user consent~\cite{ahmed2020gdpr}. Despite the GDPR being meant for European citizens, OSN services not located in the EU must still abide by the GDPR if they want to operate in the EU.}} This aspect was touched by~\citeauthor{kotsios2019analysis} in 2019, i.e., more than five years ago. Hence, we seek to investigate whether the enactment of the GDPR (in 2018) led to some ``changes'' in the trends pursued by researchers in different areas of the World---and, specifically, in four major geographical areas: the US, China, India, and the European Economic Area~(EEA).
For \texttt{RQ4} we consider two sub-RQ:

\begin{enumerate}[label=\texttt{RQ4.\alph*:}, leftmargin=2em, align=left]
    \item \textit{Regional publication trends}, where we explore whether there has been any shift in publications among researchers from the four areas of choice.

    \item \textit{Regional OSN diversity}, where we analyze potential changes in the OSN studied by researchers from in the four areas of choice.
    

\end{enumerate}

}
\paragraph{Methodology.}
\revision{For \texttt{RQ4.a}, we had extracted the country of the authors' affiliations (in \texttt{RQ2.a}): we re-use this information as-is for authors from the US, China and India; for the EEA we simply aggregate data of authors from EEA countries.
For \texttt{RQ4.b}, we compute the total occurrences of \textit{different} OSN by year for each group of authors from the four considered regions. Additionally, we compute the number of publications among the chosen groups for four popular OSN in 2024: Twitter, Facebook, Reddit and TikTok.}





\paragraph{Results.}

\begin{figure}[!t]
    \centering
    \subfigure{\includegraphics[width=0.23\textwidth]{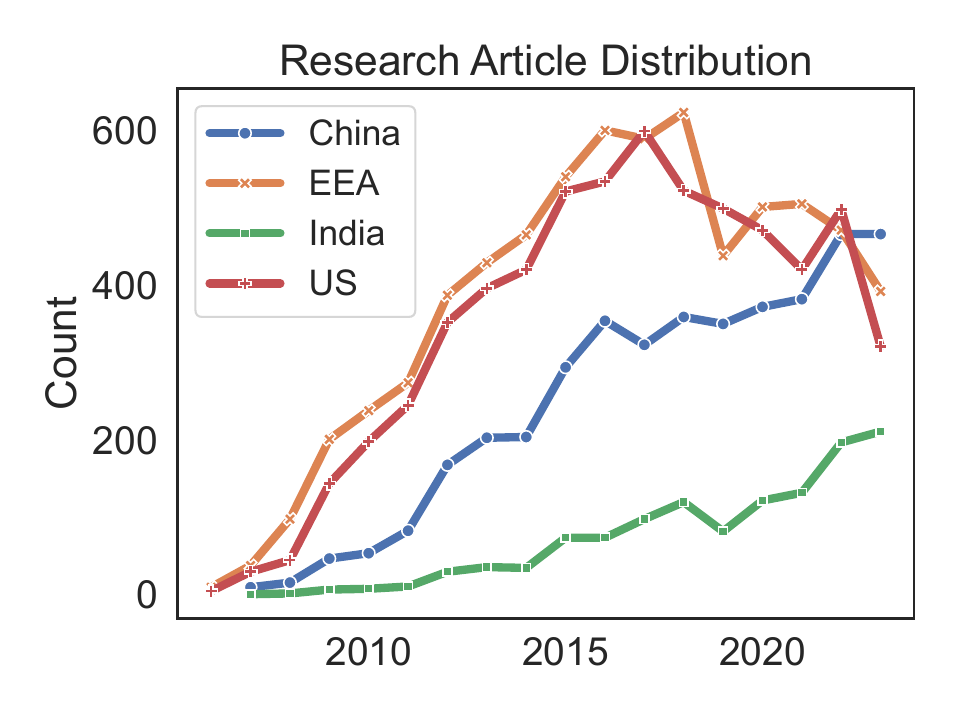}} 
    \subfigure{\includegraphics[width=0.23\textwidth]{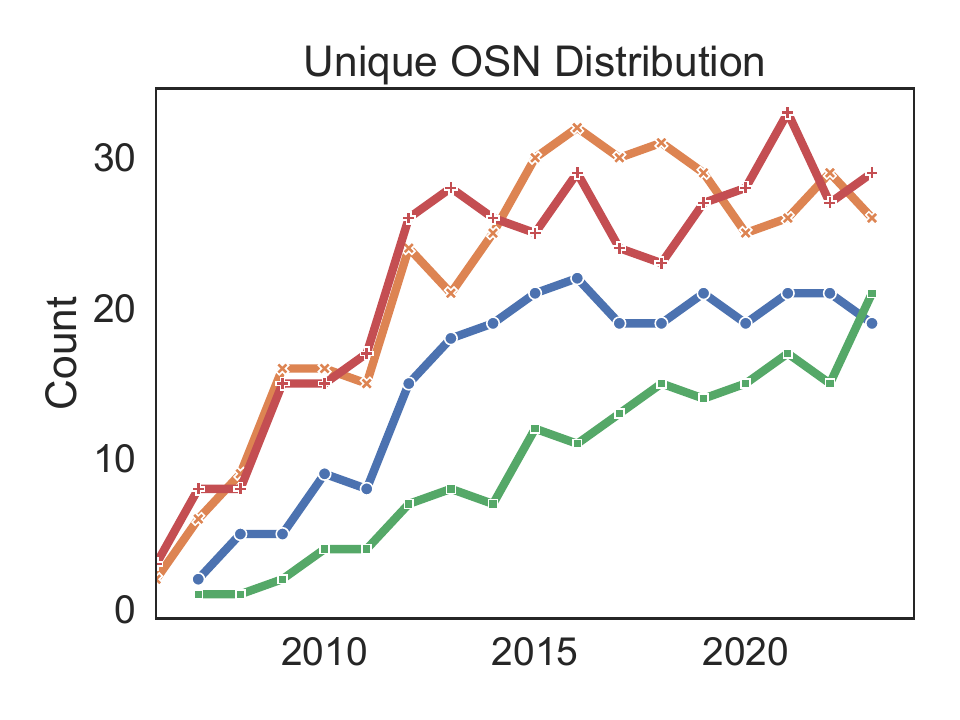}}
    \caption{\revision{Left: papers from EEA/non-EEA researchers. Right: the distribution of unique OSN over the years.}}
    \label{fig:GDPR-plots}
\end{figure}

\begin{figure*}[t]
    \centering
    \subfigure{\includegraphics[width=0.24\textwidth]{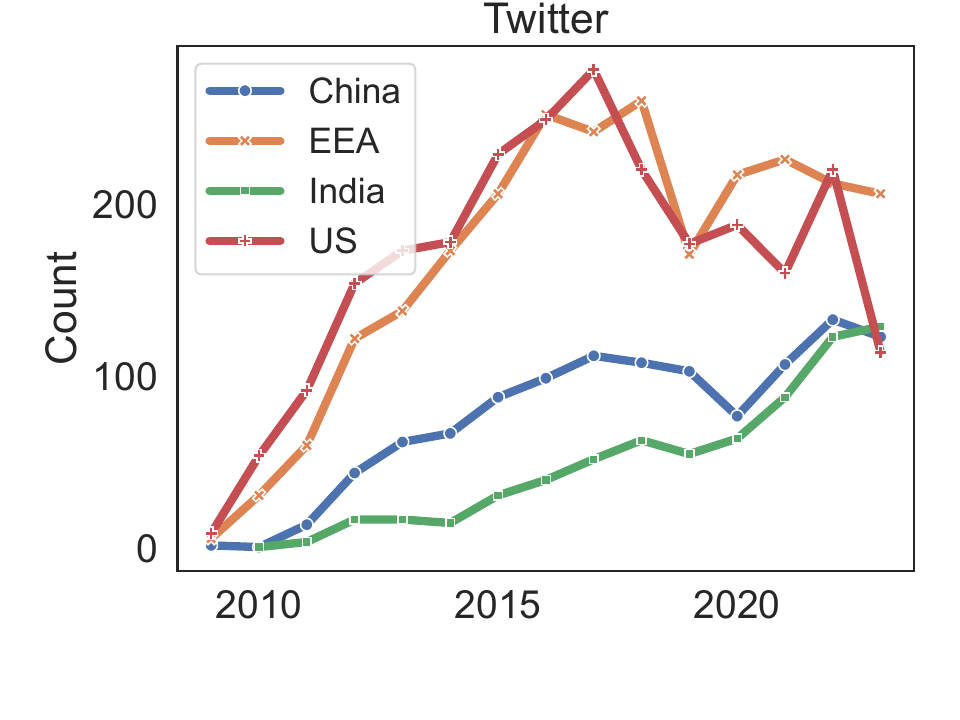}} 
    \subfigure{\includegraphics[width=0.24\textwidth]{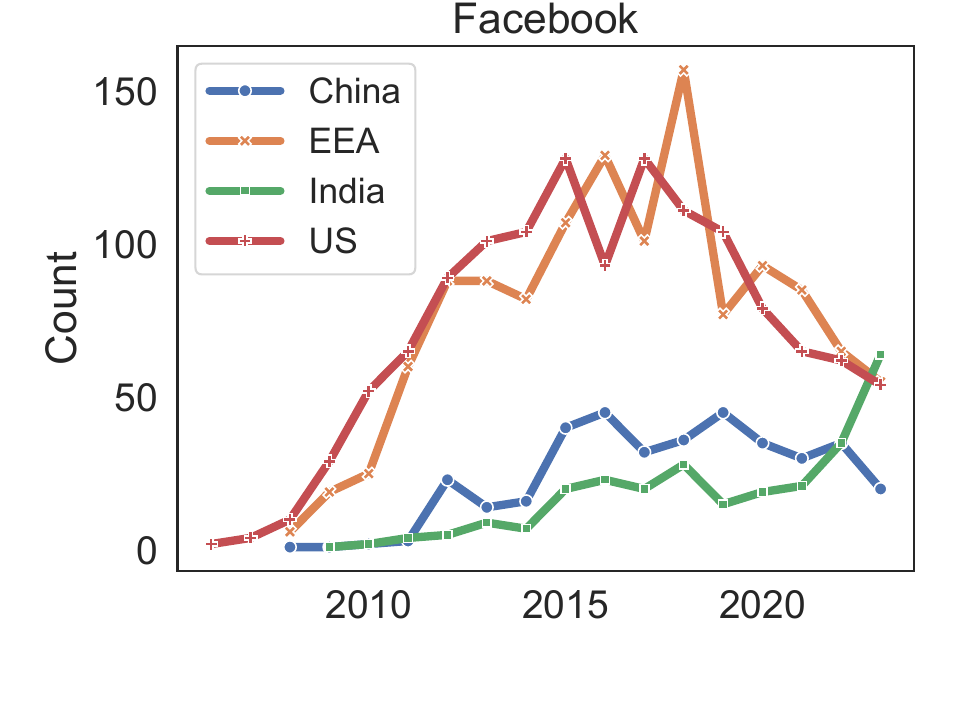}}
    \subfigure{\includegraphics[width=0.24\textwidth]{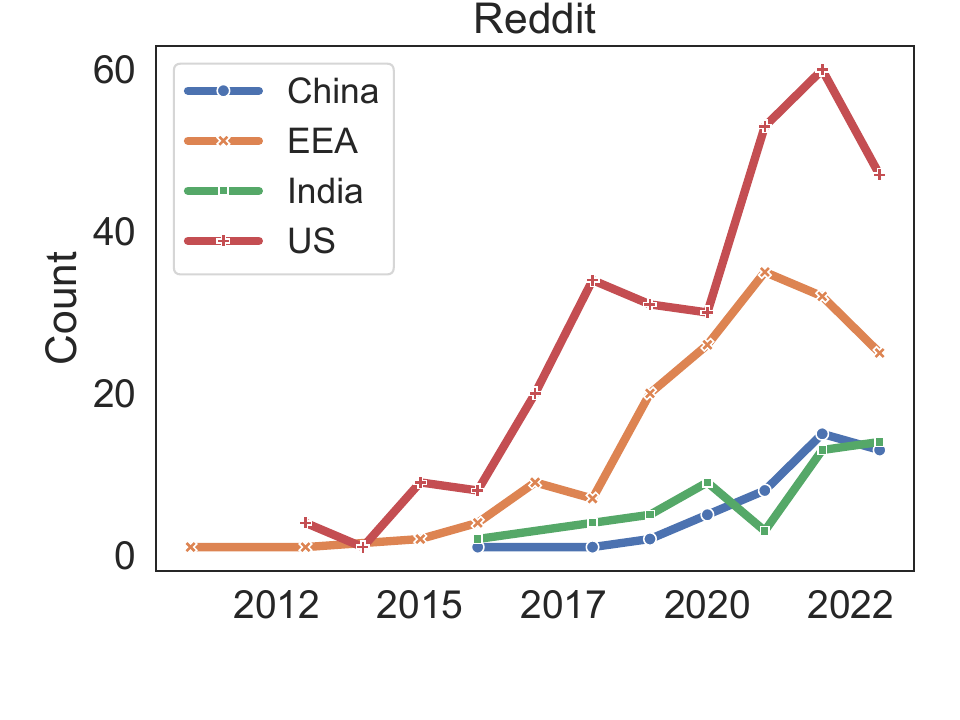} }
    \subfigure{\includegraphics[width=0.24\textwidth]{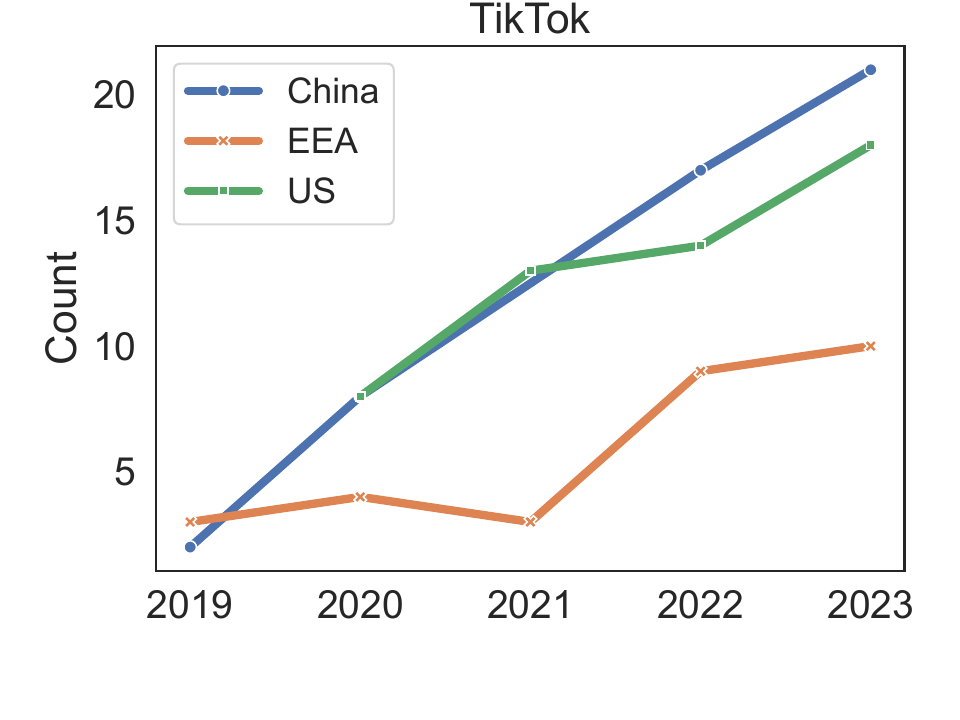}}
    \caption{\revision{\textbf{Papers (related to Twitter, Facebook, Reddit, TikTok) authored by EEA and non-EEA researchers over time.} The y-axis of each subfigure has a different scale because we compare trends between EEA and other countries for each OSN (we do not aim at carrying out cross-platform analyses). TikTok was released in 2018 so there is no data before that year.}}
    \label{fig:changes-4-OSN}
\end{figure*}

[\texttt{RQ4.a}] \revision{Figure~\ref{fig:GDPR-plots} (left) visualizes the distribution of research papers over years for authors from the four considered geographical areas. Overall, we can observe a stable rise of publications for each group---until 2017. Indeed, after 2017, we see two contrasting trends: researchers from the EEA and US have a stark drop in the number of OSN-related publications; in contrast, the OSN-research output of researchers from India or China continues to rise.}

[\texttt{RQ4.b}] 
\revision{
We plot the distribution of the \textit{different} OSN studied every year by researchers from the four considered geographical areas in Figure~\ref{fig:GDPR-plots} \revision{(right)}. Researchers from India and China study a smaller number of OSN w.r.t. researchers from the EEA or the US; we  observe a small decline of OSN diversity for EEA researchers in 2018--2019. Figure~\ref{fig:changes-4-OSN} visualizes the yearly changes for Twitter, Facebook, Reddit, and TikTok. We observe that research on TikTok is scarce (and is mostly led by China) and that there has been a decline in Facebook and Twitter after 2018. For Reddit, instead, the number of publications is increasing.

\paragraph{Discussion.}
We conjecture that the findings of \texttt{RQ3} may be due to the GDPR (effective since 2018), which led to more restrictive data-access policies from OSN-owners---which made certain platforms ``harder to study'' by many researchers from the EEA or the US. Notably, researchers from the EEA and the US were also those with the highest research output: we posit the reason why OSN-related publications from China or India did not decrease is because of a different ``ecosystem'' (as also evidenced by, e.g., a smaller diversity of studied OSN) that was not significantly affected by the GDPR. However, these are just hypotheses: there can be other reasons behind the decrease in publications from the EEA/US (e.g., maybe researchers simply turned their attention to other domains). Nonetheless, the fact 
that the GDPR influenced research (not necessarily on OSN) is recognized~\citep{greene2019adjusting,siegert2020personal}.}

\par



\begin{greentextbox}
{\scriptsize \faChartLine } \textbf{\texttt{RQ4}. How does the OSN-research output differ across geographical areas?} Starting from 2018, researchers from the US and the EEA published less papers on OSN. China and India study less OSN w.r.t. the US and EEA. Researchers from the EEA and the US have similar trends for Twitter and Facebook; however, few researchers from the EEA (and none from India) study TikTok.
\end{greentextbox}


\subsection{\texttt{RQ5}. OSN Popularity: Users vs. Researchers}
\emph{How does the popularity of OSN (among end-users) relate to their prevalence in research?} The goal of \texttt{RQ5} (which has no sub-questions) is understanding whether some OSN are over- or under-studied w.r.t. their popularity among users---thereby reflecting their importance in the real world.

\paragraph{Methodology.}
 To carry out a meaningful and feasible analysis, we restrict our investigation to the 20 most prevalent OSN in our \dataset. We then proceed as follows: \textbf{(1)}~We take a snapshot of the current OSN popularity based on its usage (as of April 2024). We refer to two sources: \textit{website ranking}~\ref{link:cloudflare-ranking}, and \textit{monthly active users}. The website ranking reflects the number of unique users accessing a domain over a specific period of time, and allows conclusions about the overall popularity of a certain OSN compared to other domains. Monthly active users is a platform-specific metric about the registered users actively using the OSN on a monthly basis. 
 \textbf{(2)}~We capture a historical snapshot of OSN popularity from 2009--2022 based on website ranking to identify how OSN popularity has changed over time by using Alexa's top website list~\ref{link:alexa-top500}. For comparison, we compute the number of published papers per OSN for 2006--2023. We capture the historical data for \revision{6} OSN. We exclude TikTok since no historical data is available. 


\begin{table}[!t]
\centering
\scriptsize
\begin{tabular}{p{1.5cm}|p{.62cm}|p{1.9cm}|p{.9cm}|p{.9cm}}
\toprule
\textbf{OSN} & \textbf{Birth} & \textbf{\# papers ($>$2016)} & \textbf{Rank} & \textbf{MAU} \\
\midrule
Twitter   & 2006 & \revision{5,248 (3,672)}  & \textbf{42}  & 611~\ref{link:statista-mau} \\
Facebook  & 2004 & \revision{2,545 (1,598)}  & \textbf{6}   & \textbf{3,065}~\ref{link:statista-mau} \\
Wikipedia & 2001 & \revision{2,139 (1,006)}  & \textbf{71}  & 0.13~\ref{link:wikipedia-mau} \\
\revision{YouTube} & \revision{2005} & \revision{1,438 (960)}  &\revision{\textbf{14}} &\textbf{2,000}~\ref{link:youtube} \\
Flickr   & 2004 & \revision{985 (350)}  & $>$1000               & 60~\ref{link:flickr-mau} \\
\revision{Weibo} & \revision{2009} & \revision{573 (438)} & \revision{$>$1000} & \revision{582~\ref{link:weibo}}\\
Instagram & 2010 & \revision{482 (445)}  & \textbf{16}     & \textbf{2,000}~\ref{link:statista-mau} \\
Reddit    & 2005 & \revision{358 (337)}  & Top1000         & \textbf{1,212}~\ref{link:reddit-mau} \\
Yelp      & 2004 & \revision{319 (268)}  & $>$1000               & 178~\ref{link:yelp-mau} \\
Moodle    & 2002 & \revision{269 (175)}  & $>$1000               & - \\
\revision{Foursquare} & \revision{2009} & \revision{255 (158)} & $>$1000 & - \\
LinkedIn  & 2003 & \revision{253 (159)} & \textbf{67}      & 310~\ref{link:linkedin} \\
\revision{StackOverflow} & \revision{2008} & \revision{159 (144)} & \revision{$>$1000} & \revision{100~\ref{link:stack}}\\
\revision{Second Life}  & \revision{2003} & \revision{119 (11)} & \revision{$>$1000} & \revision{-} \\
Google+   & 2011 & \revision{89 (40)}  & $>$1000                & - \\

Last.fm   & 2002 & \revision{85 (32)} & $>$1000                & - \\
TikTok    & 2016 & \revision{82 (82)}  & \textbf{8}       & \textbf{1,582}~\ref{link:statista-mau} \\
\revision{Telegram} & \revision{2013} & \revision{80 (79)} & \revision{Top1000} & \revision{800~\ref{link:telegram}}\\
Quora     & 2010 & \revision{78 (66)}  & Top1000        & 400~\ref{link:quora-mau} \\
\bottomrule
\end{tabular}%

\caption{\textbf{OSN prevalence in research and real-world popularity.} We report the number of papers of the 20 most prevalent OSN across the entire \dataset, and the number of papers published since the release of the youngest OSN (TikTok, in 2016); the rank of Cloudflare's global website ranking for each OSN~\ref{link:cloudflare-ranking}; the number of monthly active users (in millions); a ``-'' denotes that we could not find any reliable/exhaustive source. We highlight (in boldface) ranks below the top 100 and active users if above 1 billion.}
\label{tab:current-osn-popularity}
\end{table}

\paragraph{Results.}

We report the snapshot of OSN popularity in Table~\ref{tab:current-osn-popularity}. Facebook is still very popular in 2024, and TikTok is also among the most popular OSN. Regarding monthly active users (MAU), four OSN have more than a billion active users: Facebook, Instagram, Reddit and TikTok. Considering that only \revision{82} papers have been published about TikTok, it seems understudied compared to its current impact.

 
We report the historical analysis of OSN popularity in Figure~\ref{fig:OSN-popularity-over-time}; the x-axis ends in 2022 because we did not find 2023 data for some OSN. 
Interestingly, Twitter has been the most studied OSN in research papers, while its popularity based on website ranking is constantly decreasing. Conversely, while Facebook's popularity has been generally stable (i.e., it has been one of the top-ranked domains globally over many years), we observe a declining trend in research.
\par
Altogether, these results reveal an alarming trend: \textit{OSN research is not representative of the real world anymore}. 
Consider the following: in 2023, among people aged 18--29, the most popular OSN were BeReal, Instagram, Snapchat, and TikTok~\ref{statistaUSSocial}. These OSN are not well covered in the literature. 
Therefore, research on OSN can present \textit{nonrepresentative bias}, e.g., Twitter and Facebook are not the most suitable OSN to conduct studies in younger generations, such as on cyberbullying~\citep{whittaker2015cyberbullying}.


\begin{greentextbox}
{\scriptsize \faChartLine } \textbf{\texttt{RQ5}. How does the popularity of OSN (among end-users) relate to their prevalence in research?} 
The existing body of OSN research does not accurately reflect the dynamics of the real world. Studies utilizing Twitter as a primary data source are disproportionately high, notwithstanding its observed decrease in popularity in recent years. Consequently, research topics pertaining to younger demographics may be inadvertently overlooked.
\end{greentextbox}

\section{Data-Access Policies}
\label{sec:policies}
We attempt to explain some of the trends that we brought to light by analysing the ``data-access policies'' that each OSN enacts. In particular, we focus on policies related to the access to data pertaining to the OSN by end-users. 

\textbf{Motivation.} One of the main goals of research is advancing our knowledge by analysing factual evidence---i.e., data. In the context of OSN, end-users typically have access only to a small subset\footnote{E.g., a given Facebook user can see the posts shared by their friends and which appear on their news feed; but can hardly observe the activities performed by complete strangers.} of the whole data gravitating around the OSN. Hence, from the perspective of a researcher who wants to derive new ``general'' insights from a given OSN, or who wants to test the general effectiveness of a new method, it is impossible to carry out similar analyses if the data they have access to only pertains to a single (or a restricted subset of) user. It then follows that OSN that provide easy access to their data (e.g., via custom APIs, or public and up-to-date datasets) are more appealing to researchers---thereby explaining why more papers focus on such \revision{OSN~\citep[e.g.,][]{antonakaki2021survey}.} 

\textbf{Method.} We follow an approach similar to the one by~\citet{fiesler2020no}. Specifically, in December 2023, we manually review the policies of \revision{eight} major OSN: Facebook, Reddit, Snapchat, StackExchange, Steam, TikTok, Twitter, \revision{YouTube}. We scrutinize whether these OSN (i)~\textit{allow end-users to retrieve data} from the OSN in a ``legitimate'' way, i.e., by using APIs provided by the OSN owners---and not, e.g., via scraping or creating fake accounts; we also review (ii)~\textit{what data} can be retrieved. Finally, we scrutinize (iii)~\textit{milestones that changed} the policies of a given OSN. To provide long-term evidence of our search, we have saved the web pages documenting our statements in permanent links (reported in the appendix).

\textbf{Results.} Let us report our findings for each OSN:
\begin{itemize}
    \item \textit{Facebook} provides an API~\ref{facebook-api}, which is free (albeit it has a limit of 200 queries per hour) but it only allows to retrieve information pertaining to the account owner. 

    \item \textit{Reddit} provides an API~\ref{reddit-api} which has a limit of 1,000 queries per minute for the free version. An important change on Reddit occurred in July 2023~\ref{reddit-api2}: before this date, there was no limit to the queries that could be performed through this API. To provide some context, a popular app would require over 1.7M USD per month to sustain its traffic with the current pricing~\ref{reddit-api3}.

    \item \textit{Snapchat} provides an API to create posts~\ref{snapchat-api}, but it cannot be used to retrieve data from the OSN for research.

    \item \textit{StackExchange} provides an API~\ref{stackexchange-api} which allows free access to its data but with a limit of 10k requests per day.

    \item \textit{Steam} provides a free API~\ref{steam-api} which allows to retrieve of information about news/stats on a given video game; however, it does not allow to retrieve of data pertaining to social interactions among users (e.g., posts on boards).

    \item \textit{TikTok} provides a free API~\ref{tiktok-api} usable to customize a user's own posts (similar to SnapChat's). 

    \item \textit{Twitter} provides an API to read the content posted on its platform~\ref{twitter-api}, but it is not free: the basic version costs 100 USD per month, and allows to read at most 10k posts; whereas the pro version costs 5,000 USD per month and allows 1M posts per month. A significant change in Twitter's API occurred on March 30th, 2023: before this date, there was a free API that allowed to perform read requests (around 100 every 15 minutes).

    \item \revision{\textit{YouTube} has a free API~\ref{youtube-api} (with limited quotas) to read data from specific channels/videos or for search queries.}
\end{itemize}

\textbf{Considerations.} It is apparent that these popular OSN hardly enable ``independent'' researchers to analyzee the data posted on their platform---at least today. Indeed, some OSN, such as Twitter and Reddit, were more open to research activities---but this changed in 2023. There are, however, some OSN that do provide APIs for exclusive use to researchers. Yet, obtaining access to such APIs is not straightforward. We provide some examples of such ``difficulties'':
\begin{itemize}
    \item for Tiktok, one must submit an application form, for which a response is provided ``within 3--4 weeks''~\ref{tiktok-api-lim};
    \item for Facebook, it is also required to ``sign up to learn more Information about Facebook Open Research and Transparency including access and eligibility''~\ref{fb-api-lim};
    \item following Reddit's policy change, it was stated that unlimited usage of its API would be given to researchers: yet, as of March 2024, it is still unclear how to do so~\ref{reddit-api-lim}. 
\end{itemize} 
Put simply, obtaining access to the data of OSN for research purposes is tough for researchers, given the current policies---a finding which echoes the one by~\citet{fiesler2020no}. \textit{We acknowledge, however, that such policies can be well founded.} OSN have the data of millions of individuals, and it is reasonable that OSN owners want to ensure that only  ``trustworthy'' parties have access to their data; moreover, accessing such data has a cost, and excessive queries may put a lot of stress on the OSN owners' servers. \revision{Finally, it is also within the OSN owners' rights not to give away ``for free'' high-value data that can be monetized~\ref{link:reddit-ai}.}

\section{Researchers' Opinion}\label{sec.survey}

\textit{What do researchers think about OSN research?}
Using \dataset{} as a starting point, we contact other scientists to grasp their opinions on the state of OSN research.

\subsection{Methodology}
\paragraph{Participant Selection.}
Due to privacy restrictions in Scopus, authors' contact information are not readily available (which is why it is not included in our \dataset).
Hence, we manually retrieve contact details. To make our analysis feasible and meaningful, we first identify researchers that co-authored \textit{at least 5 papers} in our \dataset: this criterion was met by 1,138 authors. We then analyze the most recent publications of these authors to pinpoint the most plausible contact information---which we find after manually reviewing 788 papers. Additionally, recognizing the value of diverse perspectives, we include emails of co-authors of these 788 papers, given that contact authors often differ from the primary author of interest. This approach facilitates broader participation and enriches the survey with insights from younger researchers. In total, we contacted 2,367 authors via an anonymous online questionnaire. 

\paragraph{Questionnaire.} Our questionnaire is \textit{anonymous}, but we ask for some basic (and optional) information about each participant, such as age, country, and employment status; we also inquire about the participant's research background, such as which OSN they work on, years of experience in OSN research, and if they are still actively researching this field. We report a copy of the survey in our repository. The questionnaire has the following structure:
\begin{itemize}
    \item \textit{Data Gathering: Ideal vs Real World.} We ask authors to rate (1=less relevant to 5=more relevant) what option they would prefer when gathering data for OSN research. We specifically mention using open-source datasets, collecting new datasets through official APIs, collecting datasets through methods not involving APIs, and entering contact with OSN organizations. Importantly, for this part, we first ask participants to provide their answer by assuming an ``ideal world'' (i.e., with no constraints or unlimited resources), and then ask the same question but by considering the implications of the real world.

    \item \textit{Partnerships with OSN.} We ask closed questions about the authors' perceptions of the difficulty in collaborating with OSN providers and whether a more straightforward process to (i) partner up with OSN providers and (ii) obtain OSN data would facilitate future research.
    
    \item \textit{Reproducibility and Outlook}
    We ask closed questions about the authors' perception of the degree of reproducibility of prior research on OSN. The questionnaire ends with an open question in which we ask for comments and feedback on the current state of OSN research.
\end{itemize}

\paragraph{Sample Description and Limitations.}
We sent emails to 2,367 authors in May 2024 (we only sent one email to avoid annoying fellow researchers) and collected responses for one week.
In total, we received 50 answers (a participation rate of 2\%). We acknowledge our sample is small (we do not claim generality) but their opinion is still useful in driving future research. Our participants are
mainly from Europe (64\%), North America (18\%), and Asia (12\%); such a distribution may create bias (e.g., Europeans are more affected by GDPR) in their responses. Most respondents are associate professors or above (74\%), stressing that the answers we received come from experts in the field. Indeed, 86\% of participants have been doing research for more than 5 years, while only 4\% is relatively new to the field, with less than 2 years of experience (but their opinion is still scientifically valid). Most have worked on Twitter (86\%) and Facebook (72\%), followed by Wikipedia, Instagram, and Reddit (around 35\% each), which partially reflects the popularity of such OSN in general research (and may, hence, cause some bias). 74\% of participants are still doing research on OSN.




\subsection{Results}
 

\begin{figure}
    \centering
    \includegraphics[width=\linewidth]{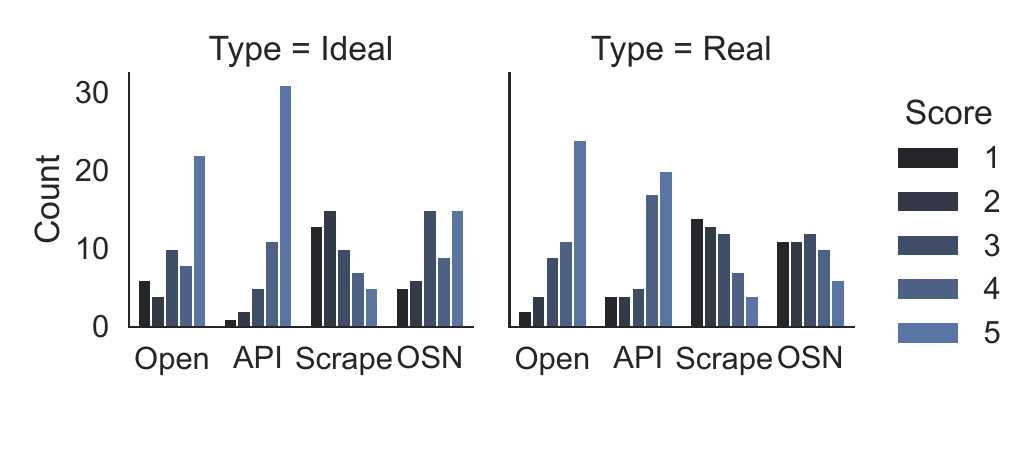}
    \caption{\textbf{Researchers preference in data gathering: ideal vs. real world.} We indicate preference (1 is low, 5 is high) to rely on open source datasets (``Open''), creating a new dataset through official APIs (``API'') or by scraping (``scrape''), or via a direct channel with the OSN (``OSN'').}
    \label{fig:idealvsreal}
\end{figure}


\paragraph{Data Gathering (Ideal vs. Real-world).}
We compare ``what researchers would like to use'' (in an ideal world) against ``what researchers are forced to use'' (by accounting for the constraints of the real world). 
Figure~\ref{fig:idealvsreal} presents the comparative results for both scenarios, with a rating scale from 1 (lowest preference) to 5 (highest). Notably, alternative methods (e.g., scraping) are not among our participants' priorities. Using APIs was the preferred choice in an ideal scenario, but reusing open-source datasets is the first option in the real world. More participants consider contacting OSN owners in an ideal world w.r.t. the real world.

\paragraph{Partnerships with OSN.}
When asked if they had engaged in direct research collaborations with any OSN, 78\% of participants responded negatively, with 98\% of them confirming that such collaborations are challenging to set up (30\% consider it ``hard,'' while 50\% ``very hard''). These findings may explain why partnerships with OSN have relatively low ratings in the ideal-vs-real--world questions. As a matter of fact, 92\% of participants agreed that there should be a more straightforward process to partner up with OSN, given that it could improve the state of OSN research (96\% said so). 

\paragraph{Reproducibility and Outlook.}
In terms of reproducibility, while 22\% of participants admitted to never having conducted a reproducibility study, the majority found the process challenging. Specifically, 34\% described it as ``difficult,'' while 26\% found it ``very difficult.'' Another 12\% remained neutral on the matter. Only 2 participants considered it ``easy.'' With regards to the last (open and optional question), we report below five excerpts of the statements we received, which we find particularly inspiring:
\begin{itemize}
    \item ``Twitter closed their academic collaboration program, we lost access to any meaningful data for research. Around the same time, Reddit also created restrictions, leaving us with no data to use and almost impossible to do any research unless there is lots of money to pay.''
    \item ``OSN will work with you if you are a paid contractor.''
    \item ``I hope the government will force OSN companies to share their data with the researchers.''
    \item ``OSN should make their data available for academic purposes---research and teaching. OSN are popular media nowadays, especially among young people, and valuable insights can be extracted from them.''
    \item ``IRB and Office of Sponsored Projects tend to have a lot of legal issues and hurdles for university researchers.''
\end{itemize}

\textbox{\textsc{\textbf{Takeaway.}} OSN researchers perceive obtaining data (and reproducing prior works) as challenging tasks, and that making them easier would facilitate future research.}
\section{Discussion}
Our investigation highlights that researchers encounter substantial challenges in studying OSN---a problem that is exacerbated by the numerous restrictions that OSN owners enacted over time. 
To fix these issues, our findings should become a subject of discussion not only among stakeholders within the OSN ecosystem (i.e., researchers and OSN owners\footnote{We argue that research on OSN should transcend simplistic (but fair and legitimate) visions such as ``career trajectories of researchers'' or ``a yet another source of profit for OSN owners.''}), but also among policymakers. Given the billion-sized userbase of OSN, and the potential of OSN to influence our society, \textit{a smooth development of such platforms should be within the interest of political institutions}. According to our study, researchers have been advocating for such a change---but their ``voices'' have not been adequately heard, it appears.

Throughout the course of our study, we identified numerous indicators suggesting that the priorities of researchers may not be fully aligned with the real world. 
The most evident sign of this phenomenon can be appreciated in Figure~\ref{fig:OSN-popularity-over-time}, showing that \textit{Twitter has been the most represented OSN in research since 2011, despite being outside the top-3 OSN (popularity-wise) since 2015}. Such an asymmetry is particularly detrimental to younger demographics (who tend to use OSN that are less prevalent in research, such as Instagram and TikTok), as potential risks remain unexplored due to researchers’ propensity to focus on OSN that are ``easier to study'' -- despite the fact that such OSN may not fully capture the entire spectrum of our society. 

Let us elaborate on our previous claim. 
Consider the heatmap in Figure~\ref{fig:osn_topics}: Twitter is a key resource in 10 out of the 17 most commonly studied research topics.
Such a finding suggests that Twitter is---or rather, \textit{was}---a cornerstone for OSN research, owing to its open API and policies that permit straightforward access to its data, thereby fostering the collection and distribution of new datasets to the community. Such a characteristic, however, is not valid anymore: the easy accessibility to Twitter's data was curtailed in April 2023 with the introduction of (expensive!) paid tiers in its API. In contrast, other OSN, such as Instagram and TikTok, have always posed challenges to researchers interested in analyzing their data. We stress, however, that \textit{there is nothing wrong in carrying out research on an ``overly-studied'' OSN}. However, it is factual that some OSN have a different user base, and lack of research on such platforms is detrimental to their communities (and, hence, to our whole society).


Our findings have also highlighted a decreased ``interest'' from researchers towards OSN---despite the skyrocketing growth of these platforms' userbase. Indeed, \textit{starting from 2018, less researchers has been attracted in OSN research, and less peer-reviewed papers on OSN have appeared} (see \texttt{RQ1.a}, 
 \texttt{RQ2.a}, \texttt{RQ4}). This can be linked to the enactment of country-wide policies such as the GDPR, as well as by -- potentially -- a reduced attractiveness of ``overused'' OSN to researchers (which may have also negatively affected the desirability of certain submissions to some venues). Hence, the question arises: how will the recent ``restrictive'' policy changes on well-researched OSN (such as Reddit and Twitter) affect future research on OSN?\footnote{An intriguing initiative that seeks to mitigate the poor access to OSN data is Perspective~\ref{link:perspectiveAPI}, but it only pertains to toxic content.}

\par
\par
\par
%

\section{Recommendations and Future Work}

We have provided a holistic overlook of extant research on OSN by analyzing the metadata of \revision{13,842} publications---which we organized in the \dataset\ dataset, which we publicly release. \revision{Our results highlight some fundamental issues in the OSN ecosystem. To ``rescue'' OSN-related research, we make the following recommendations:}
\begin{itemize}
    \item \textit{OSN owners should ``lend a hand'' to OSN researchers}. Restrictions to OSN data impair research, which is detrimental to all OSN stakeholders.

    \item \textit{OSN researchers should explore ``other'' OSN}. Each OSN caters to a specific segment of the world's population, and investigating such platform may reveal aspects of our society that deserve to be brought to light.

    \item \textit{Policymakers (at all levels) should contribute, too.} OSN are a pivotal component of our society, and certain policies can be waived for research purposes.
\end{itemize}
Our contributions are meant to guide future work on OSN, inspiring to, e.g., explore new topics (or on new OSN); or to raise awareness of the issues faced by researchers to other relevant stakeholders. Finally, our \dataset\ dataset, codebase, \revision{and analytical methods} can be used by downstream research to investigate other aspects related to OSN.

\section*{Ethical Statement}
Our institutions are aware of the research carried out in this paper, and deemed that a formal IRB approval was not required. Our \dataset\ was composed by collecting publicly available information, hence making it public does not violate any copyright. The participants of our survey were informed of the purpose of our (anonymous) questionnaire, with the option to withdraw their data if they so desire. 

\section*{Acknowledgements}
We thank the reviewers for their comments: addressing your feedback improved our paper immensely. We want to ex tend our gratitude to all the researchers who participated in
 our study. Part of this research has been funded by the Hilti Foundation.



{\small 
\bibliography{main}
}


\section{Paper Checklist}

\begin{enumerate}

\item For most authors...
\begin{enumerate}
    \item  Would answering this research question advance science without violating social contracts, such as violating privacy norms, perpetuating unfair profiling, exacerbating the socio-economic divide, or implying disrespect to societies or cultures?
    \answerYes{Yes}
  \item Do your main claims in the abstract and introduction accurately reflect the paper's contributions and scope?
    \answerYes{Yes}
   \item Do you clarify how the proposed methodological approach is appropriate for the claims made? 
    \answerYes{Yes}
   \item Do you clarify what are possible artifacts in the data used, given population-specific distributions?
    \answerYes{Yes}
  \item Did you describe the limitations of your work?
    \answerYes{Yes}
  \item Did you discuss any potential negative societal impacts of your work?
    \answerYes{Yes, in the Ethics and Limitations.}
      \item Did you discuss any potential misuse of your work?
    \answerYes{Yes, in the Ethical Statement.}
    \item Did you describe steps taken to prevent or mitigate potential negative outcomes of the research, such as data and model documentation, data anonymization, responsible release, access control, and the reproducibility of findings?
     \answerYes{Yes}
  \item Have you read the ethics review guidelines and ensured that your paper conforms to them?
    \answerYes{Yes}
\end{enumerate}

\item Additionally, if your study involves hypotheses testing...
\begin{enumerate}
  \item Did you clearly state the assumptions underlying all theoretical results?
    \answerNA{NA}
  \item Have you provided justifications for all theoretical results?
    \answerNA{NA}
  \item Did you discuss competing hypotheses or theories that might challenge or complement your theoretical results?
    \answerNA{NA}
  \item Have you considered alternative mechanisms or explanations that might account for the same outcomes observed in your study?
    \answerNA{NA}
  \item Did you address potential biases or limitations in your theoretical framework?
    \answerNA{NA}
  \item Have you related your theoretical results to the existing literature in social science?
    \answerNA{NA}
  \item Did you discuss the implications of your theoretical results for policy, practice, or further research in the social science domain?
    \answerNA{NA}
\end{enumerate}

\item Additionally, if you are including theoretical proofs...
\begin{enumerate}
  \item Did you state the full set of assumptions of all theoretical results?
    \answerNA{NA}
	\item Did you include complete proofs of all theoretical results?
    \answerNA{NA}
\end{enumerate}

\item Additionally, if you ran machine learning experiments...
\begin{enumerate}
  \item Did you include the code, data, and instructions needed to reproduce the main experimental results (either in the supplemental material or as a URL)?
    \answerYes{Yes, we provided a URL with source code and data.}
  \item Did you specify all the training details (e.g., data splits, hyperparameters, how they were chosen)?
    \answerYes{Yes}
     \item Did you report error bars (e.g., with respect to the random seed after running experiments multiple times)?
    \answerNA{NA, as we only utilize unsupervised machine learning.}
	\item Did you include the total amount of compute and the type of resources used (e.g., type of GPUs, internal cluster, or cloud provider)?
    \answerYes{Yes, specific details are reported in our repository.}
     \item Do you justify how the proposed evaluation is sufficient and appropriate to the claims made? 
    \answerYes{Yes, the analyses we conducted are confirmed by researchers' opinion in our survey.}
     \item Do you discuss what is ``the cost`` of misclassification and fault (in)tolerance?
    \answerNA{NA, as we only utilize unsupervised machine learning.}
  
\end{enumerate}

\item Additionally, if you are using existing assets (e.g., code, data, models) or curating/releasing new assets, \textbf{without compromising anonymity}...
\begin{enumerate}
  \item If your work uses existing assets, did you cite the creators?
    \answerNA{NA, as we only utilized the dataset we collected.}
  \item Did you mention the license of the assets?
    \answerNA{NA}
  \item Did you include any new assets in the supplemental material or as a URL?
    \answerNA{NA}
  \item Did you discuss whether and how consent was obtained from people whose data you're using/curating?
    \answerYes{Yes, this information is available in the survey description.}
  \item Did you discuss whether the data you are using/curating contains personally identifiable information or offensive content?
    \answerNA{NA}
\item If you are curating or releasing new datasets, did you discuss how you intend to make your datasets FAIR?
\answerYes{Yes. The discussion can be found in the dataset section.}
\item If you are curating or releasing new datasets, did you create a Datasheet for the Dataset? 
\answerYes{Yes}
\end{enumerate}

\item Additionally, if you used crowdsourcing or conducted research with human subjects, \textbf{without compromising anonymity}...
\begin{enumerate}
  \item Did you include the full text of instructions given to participants and screenshots?
    \answerYes{Yes, in the online repository we share a copy of the survey.}
  \item Did you describe any potential participant risks, with mentions of Institutional Review Board (IRB) approvals?
    \answerYes{Yes, we explained in the ethical considerations section that IRB was not required, participants were not exposed to risks, and their privacy was preserved.}
  \item Did you include the estimated hourly wage paid to participants and the total amount spent on participant compensation?
    \answerNA{NA}
   \item Did you discuss how data is stored, shared, and deidentified?
   \answerYes{Yes}
\end{enumerate}

\end{enumerate}

\section{Appendix}
\begin{table*}[!htpb]
\tiny
    \centering
\begin{tabular}{p{23mm}p{120mm}l}
\toprule
\textbf{Topic Name} & \textbf{Description} & \textbf{Size}\\ \midrule
Multimedia Retrieval and Tagging & It focuses on developing methods to efficiently search and categorize visual content, such as images and videos, often leveraging techniques like face recognition and object detection. This research area is crucial for enhancing user experiences on OSN by enabling more accurate and relevant content discovery. & 1,426 \\ \hline
Knowledge Mining  & It is a research area focused on identifying and associating textual mentions of entities with their corresponding entries in a knowledge base, ensuring accurate integration of information from diverse sources. This field leverages semantic analysis and data integration techniques to enhance the understanding and organization of information across the web and various datasets. & 1,305\\ \hline
Security and Privacy Risks & The research area focuses on identifying and mitigating the privacy and security threats that users face on OSN. This includes protecting user information from unauthorized access, preventing data breaches, and addressing issues like spam and fake accounts. & 1,110\\ \hline
Social Network Analysis & It is a research area that uses networks and graph theory to study social structures by mapping and measuring relationships and flows among people, groups, or organizations. It helps identify influential users, information dissemination patterns, and the overall connectivity within OSN. & 1,082\\ \hline
Location-Based Analysis & It integrates social networking with location-based services, allowing users to share and access geo-related information such as locations, trajectories, and points of interest (POIs) in real-time. This research area leverages vast amounts of spatiotemporal data to analyze user behavior and interactions, offering applications in marketing, recommendations, and urban planning. & 876\\ \hline
Sentiment Analysis & It is a research area focused on identifying and classifying emotions expressed in text, particularly on OSN like Twitter. It involves using language-based models to analyze and categorize sentiments in tweets and other text data. & 756\\ \hline
Political Analysis & This research area examines how OSN influence political behavior and public opinion, particularly during elections. This research area involves studying user interactions, media consumption, and the dissemination of political information to understand and analyze the impact on public discourse and electoral outcomes. & 746\\ \hline
eLearning & This research area explores how students interact with and participate in virtual educational settings, focusing on factors like course design, teacher involvement, and the use of OSN. This research aims to enhance learning outcomes by understanding and improving the ways students engage with online courses and educational tools. & 731\\ \hline
Social Media Use and User Experience & The research area focuses on studying how people interact with OSN and the impact of these interactions on their personalities and behaviors. It examines various aspects of social media use, including user behavior, online engagement, and the psychological effects on individuals. & 698\\ \hline
Recommendation Systems & Recommendation Systems are algorithms designed to suggest relevant items to users based on their preferences and behaviors, often leveraging OSN and user interactions. These systems analyze data from users and items to provide personalized recommendations, enhancing user experience across various platforms. & 635\\ \hline
Event Detection and Topic Analysis  & This research area focuses on identifying and analyzing significant occurrences (events) from social media platforms like Twitter by examining user-generated content such as tweets, hashtags, and other data. This research area leverages techniques to detect real-time events, track topics, and analyze trends within OSN to provide insights into social dynamics and emergent phenomena. & 614\\ \hline
Public Health Analysis & This research area involves examining data from OSN like tweets to understand and address public health issues such as mental health, drug use, and depression. This research area leverages social media data to identify trends, monitor health-related behaviors, and inform public health interventions. & 333\\ \hline
Fake News Detection & This research area focuses on identifying and mitigating the spread of false information, rumors, and misinformation across social media and news platforms. It involves analyzing the credibility of information and sources to ensure the accuracy and reliability of the content shared with the public. & 318\\ \hline
Covid-19 Social Media Analysis & This research area examines how social media platforms were used to disseminate information, track public sentiment, and spread misinformation during the Covid-19 pandemic. This research area focuses on analyzing tweets and other social media content to understand public health communication, vaccine perceptions, and the impact of misinformation on public behavior. & 282\\ \hline
Question Answering & It is a field of research focused on developing systems that can automatically respond to questions posed in natural language by retrieving relevant information from a knowledge base or generating answers based on learned models. This area encompasses various techniques and models to understand, interpret, and generate human-like dialogue, aiming to provide accurate and contextually appropriate answers. & 251\\ \hline
Online Video Quality and User Experience & The research area focuses on evaluating and enhancing the quality of video streaming services to ensure optimal user satisfaction. It involves analyzing various factors such as network conditions, content delivery, and user feedback to improve the Quality of Experience (QoE) for live and on-demand video content. & 235\\ \hline
Hate Speech Detection & This research area focuses on developing algorithms and tools to identify and reduce harmful language, including hate speech and cyberbullying, on OSN. This involves analyzing text for offensive and hateful content to create safer and more inclusive digital environments. & 208\\ \hline
\end{tabular}
    \caption{Description of the topic identified in ~\dataset.}
    \label{tab:topic}
\end{table*}

\subsection{Prompts}
\paragraph{Prompt1: Topic Description}
Describe in two sentences the research area named [TOPIC NAME] and defined by the following keywords: [KEYWORDS].

\paragraph{Prompt2: Topic Labelling prompt}
[INST]
You are a helpful, respectful and honest assistant for labeling topics.
I have a topic that contains the following documents:
[DOCUMENTS]
The topic is described by the following keywords: [KEYWORDS].
Based on the information about the topic above, please create a short label of this topic. Make sure you to only return the label and nothing more.
[/INST]

\subsection{Lists}
In this section, we report the lists utilized in our analyses. 

\paragraph{Google Scholar Subcategories.}
Artificial intelligence, Computational Linguistics, Computer Security and Cryptography, Computer Vision and Pattern Recognition, Data Mining and Analyses, Databases and Information Systems, Human Computer Interaction, Multimedia.

\paragraph{Venues.}
ACM Asia Conference on Computer and Communications Security; 
ACM Conference on Recommender Systems;
ACM International Conference on Information and Knowledge Management;
ACM International Conference on Multimedia;
ACM International Conference on Multimedia Retrieval;
ACM International Conference on Web Search and Data Mining;
ACM International Conference Proceeding Series;
ACM Multimedia Systems Conference;
ACM SIGIR Conference on Research and Development in Information Retrieval
ACM SIGKDD International Conference on Knowledge Discovery \& Data Mining
ACM SIGMOD International Conference on Management of Data;
ACM Symposium on Computer and Communications Security;
ACM Symposium on User Interface Software and Technology;
ACM Transactions on Computer-Human Interaction;
ACM Transactions on Intelligent Systems and Technology;
ACM Transactions on Knowledge Discovery from Data;
ACM Transactions on Multimedia Computing, Communications, and Applications;
ACM/IEEE International Conference on Human Robot Interaction;
Annual Meeting of the Special Interest Group on Discourse and Dialogue;
Applied Intelligence;
Applied Soft Computing;
Artificial Intelligence Review;
Behaviour \& Information Technology;
Big Data Mining and Analytics;
BlackboxNLP Workshop on Analyzing and Interpreting Neural Networks for NLP;
British Machine Vision Conference;
Computational Linguistics;
Computational Visual Media;
Computer Human Interaction;
Computer Speech \& Language;
Computer Vision and Image Understanding;
Computers \& Security
Conference of the European Chapter of the Association for Computational Linguistics;
Conference of the North American Chapter of the Association for Computational Linguistics: Human Language Technologies;
Conference on Computational Natural Language Learning;
Conference on Empirical Methods in Natural Language Processing;
Conference on Robot Learning;
Data Mining and Knowledge Discovery;
Engineering Applications of Artificial Intelligence;
EURASIP Journal on Image and Video Processing;
Expert Systems with Applications;
IACR Transactions on Cryptographic Hardware and Embedded Systems;
IEEE Conference on Multimedia Information Processing and Retrieval;
IEEE European Symposium on Security and Privacy;
IEEE International Conference on Advanced Video and Signal-Based Surveillance;
IEEE International Conference on Automatic Face \& Gesture Recognition;
IEEE International Conference on Big Data;
IEEE International Conference on Data Mining;
IEEE International Conference on Image Processing;
IEEE International Conference on Multimedia and Expo;
IEEE International Conference on Trust, Security and Privacy in Computing and Communications;
IEEE Security \& Privacy;
IEEE Spoken Language Technology Workshop;
IEEE Symposium on Security and Privacy;
IEEE Transactions on Affective Computing;
IEEE Transactions on Big Data;
IEEE Transactions on Circuits and Systems for Video Technology;
IEEE Transactions on Dependable and Secure Computing;
IEEE Transactions on Fuzzy Systems;
IEEE Transactions on Human-Machine Systems;
IEEE Transactions on Image Processing;
IEEE Transactions on Information Forensics and Security;
IEEE Transactions on Knowledge and Data Engineering;
IEEE Transactions on Multimedia;
IEEE Transactions on Neural Networks and Learning Systems;
IEEE Transactions on Pattern Analysis and Machine Intelligence;
IEEE Transactions On Systems, Man And Cybernetics Part B, Cybernetics;
IEEE/CVF Conference on Computer Vision and Pattern Recognition;
IEEE/CVF International Conference on Computer Vision;
IEEE/CVF International Conference on Computer Vision Workshops;
IEEE/CVF Winter Conference on Applications of Computer Vision;
IET Image Processing;
Information Processing \& Management;
Information Systems;
International Conference on 3D Vision;
International Conference on Advances in Social Networks Analysis and Mining;
International Conference on Artificial Intelligence and Statistics;
International Conference on Computational Linguistics;
International Conference on Data Engineering;
International Conference on Intelligent User Interfaces;
International Conference on Language Resources and Evaluation;
International Conference on Learning Representations;
International Conference on Machine Learning;
International Conference on Natural Language Generation;
International Conference on Pattern Recognition;
International Conference on Web and Social Media;
International Joint Conference on Artificial Intelligence;
International Journal of Child-Computer Interaction;
International Journal of Computer Vision;
International Journal of Data Science and Analytics;
International Journal of Human Computer Studies;
International Journal of Human-Computer Interaction;
International Journal of Interactive Mobile Technologies;
International Society for Music Information Retrieval Conference;
International Workshop on Quality of Multimedia Experience;
International Workshop on Semantic Evaluation;
International World Wide Web Conferences;
Journal of Big Data;
Journal of Information Security and Applications;
Journal of Visual Communication and Image Representation;
Knowledge and Information Systems;
Knowledge-Based Systems;
Language Resources and Evaluation;
Lecture Notes in Computer Science;
Medical Image Analysis;
Meeting of the Association for Computational Linguistics;
Multimedia Systems;
Multimedia Tools and Applications;
Multimodal Technologies and Interaction;
Neural Computing and Applications;
Neural Information Processing Systems;
Neural Networks;
Neurocomputing;
Online Social Networks and Media;
Pattern Recognition;
Pattern Recognition Letters;
Proceedings of the ACM on Human-Computer Interaction;
Proceedings of the ACM on Interactive, Mobile, Wearable and Ubiquitous Technologies;
Security and Communication Networks;
Semantic Web;
SIAM International Conference on Data Mining;
Signal Processing: Image Communication;
Social Network Analysis and Mining;
Symposium On Usable Privacy and Security;
The Journal of Machine Learning Research;
Transactions of the Association for Computational Linguistics;
Transactions on Asian and Low-Resource Language Information Processing;
Universal Access in the Information Society;
USENIX Security Symposium;
Virtual Reality;
Wiley Interdisciplinary Reviews: Data Mining and Knowledge Discovery;
Workshop on Innovative Use of NLP for Building Educational Applications;
Workshop on Machine Translation;
Workshop on Representation Learning for NLP;
World Wide Web.

\paragraph{OSN.}
We now report information about the OSN we analyzed accompanied by their number of mentions in OSN research abstracts. 
\begin{itemize}
    \item \textit{Widely studied OSN (at least 100 mentions)}: 
    \revision{Twitter (5248), Facebook (2545), Wikipedia (2139), YouTube (1438), Flickr (985), Sina Weibo (573), Instagram (482), Reddit (358), Yelp (319), Moodle (269), Foursquare City Guide (255), LinkedIn (253), Stack Overflow (159), Second Life (119).}
    \item \textit{Frequently studied OSN (between 20 and 100)}: 
    \revision{Google+ (89), Last.fm (85), TikTok (82), Telegram (80), Quora (78), douban (74), Pinterest (64), Myspace (63), Delicious (55), Tumblr (47), Meetup (46), WordPress (45), Snapchat (39), LiveJournal (33), ResearchGate (29), Flixster (23), LibraryThing (23).}
    \item \textit{Rarely studied OSN (less than 20 mentions):} 
    \revision{Goodreads (19), Duolingo (17), Discord (16), VK (16), Gab (14), Renren (12), 4chan (12), XING (12), ICQ (7), StumbleUpon (7), CaringBridge (6), Periscope (6), Yammer (6), Mastodon (6), CouchSurfing (6), Diaspora* (6), Parler (6), PatientsLikeMe (5), Steemit (5), Blogger (5), Plurk (5), Grindr (5), mixi (5), hi5 (4), Tuenti (4), Skyrock (4), aNobii (4), SoundCloud (4), Steam (3), Habbo (3), Ask.fm (3), DailyStrength (3), Cyworld (3), Koo (2), Academia.edu (2), Fotolog (2), BitChute (2), Chictopia (2), Ning (2), Buzznet (2), FilmAffinity (2), Friendica (2), Qzone (1), Odnoklassniki (1), Voat (1), eToro (1), Identi.ca (1), Blind (1), StudiVZ (1), DLive (1), Inspire (1), Xanga (1), Foursquare Swarm (1), Epik (1), Bebo (1), 8kun (1), Fetlife (1), About.me (1), Tagged (1), Fitocracy (1)}. 
    \item \textit{Never studied OSN}: \revision{
    23snaps, 8tracks.com, Amikumu, Anphabe.com, AsianAve.com, Athlinks, Band, BeReal, Biip.no, BitClout, BlackPlanet, Brainly, Busuu, CafeMom, Care2, Cellufun, Chess.com, Classmates.com, Cloob, CloutHub, CozyCot, CrossFit, Crunchyroll, Cucumbertown, DTube, DXY.cn, Dayviews, Dead Runners Society, DeviantArt, Dol2day, Doximity, Draugiem.lv, Dreamwidth, Dulwich OnView, Elixio, Ello, English, baby!, Experts Exchange, Faces, Fark, FictionCity, Fieldoo, Fillos de Galicia, Filmow, Fishbrain, Flickchart, Focus.com, Fotki, Frank, Fyuse, Gaia Online, GameFAQs, GameTZ.com, Gapo, Gapyear.com, Gays.com, Gaysir, Geni.com, Gentlemint, Gettr, GirlsAskGuys, GiveSendGo, Glee.com, GovLoop, HR.com, Hello, Hospitality Club, Houseparty, Hub Culture, I Had Cancer, IMVU, IRC-Galleria, Ibibo, Indaba Music, Influenster, Infogalactic, ItsMy/GameCloud, JamiiForums, JustPaste.it, Kaixin001, KakaoStory, Kiwibox, KizlarSoruyor, Kobo, Koofers, Kroogi, Labroots, LambdaMOO, Letterboxd, LifeKnot, Likee, LimeWire, LinguaLeo, LinkExpats, Listography, Lunchclub, MEETin, Marco Polo, MeWe, Metapedia, Miaopai, Micro.blog, Minds, MocoSpace, MouthShut.com, Mubi, My World@Mail.Ru, MyHeritage, MyLife, NK.pl, Nasza-klasa.pl, Nearby, Newgrounds, Nexopia, Nextdoor, Odysee, Open Diary, PEERtrainer, Partyflock, Patriots.win, Peach, PewTube, Pink Petro, Pixnet, Plaxo, Play.fm, Playlist.com, Poa.st or Poast, Portfolium, Postcrossing, Quechup, RallyPoint, Raptr, RateItAll, Ravelry, Readgeek, Reverbnation, Rooster Teeth, Rumble, Ryze, Sarahah, Sgrouples, Sharesome, Skoob, Snow, Solaborate, Something Awful, Soup.io, Spacehey, Spaces, Spot.IM, Spoutible, Stage 32, SubscribeStar, TV Time, TV Tropes, TakingITGlobal, Tal Canal, Talenthouse, TalkBizNow, Taringa!, TermWiki, The Meta Network, The Sphere, The Student Room, The WELL, Thinkspot, Threads, Total Recut, TravBuddy.com, Travellerspoint, Triller, Trombi, Truth Social, Twoo.com (netlog), Untappd, Vampirefreaks.com, Vero, Viadeo, Vingle, Virb, WAYN, WT Social, Warm Showers, Wattpad, We Heart It, Werkenntwen, Whisper, WikiWikiWeb, Wooxie, Woozworld, WriteAPrisoner.com, Wykop.pl, Xt3, YTMND, Yo, aSmallWorld, alimero, beBee, iNaturalist, italki, zoo.gr, İnci Sözlük.
    }. 
\end{itemize}

\revision{
\paragraph{Top50 authors' countries (in the entire \dataset{}).} 
United States (6220), China (4151), India (1240), Germany (1079), United Kingdom (1036), Japan (892), Italy (841), Australia (789), Spain (714), France (616), Canada (604), Singapore (571), Brazil (487), Taiwan (469), South Korea (449), Netherlands (391), Greece (368), Hong Kong (358), Portugal (277), Switzerland (249), Austria (240), Ireland (196), Saudi Arabia (187), Turkey (178), Russian Federation (167), Iran (159), Finland (156), Israel (148), Poland (145), Malaysia (143), Norway (138), Indonesia (137), Tunisia (122), Mexico (120), Pakistan (120), Viet Nam (116), Qatar (114), New Zealand (101), Belgium (99), Denmark (97), Sweden (96), Thailand (87), United Arab Emirates (75), Egypt (71), Chile (61), Czech Republic (59), Argentina (53), Philippines (53), Romania (52), Bangladesh (47).
}
\subsection{List of External Links}
\label{app:links}
We provide the full list of links referenced in the paper:

\begin{enumerate}[label={[\arabic*]}]
    \item \url{https://anonymous.4open.science/r/Minerva-OSN-3841}\label{link:repository}
    \item \url{https://huggingface.co/spaces/mteb/leaderboard}\label{link:Huggingface leaderboard}
    \item \url{https://blog.flickr.net/en/2018/11/01/changing-flickr-free-accounts-1000-photos/}\label{link:flickr-1TB}
    \item \url{https://www.linkedin.com/pulse/important-linkedin-statistics-data-trends-oleksii-bondar-pqlie/}\label{link:linkedin}
    \item \url{https://web.archive.org/web/20231130194404/https://developers.facebook.com/docs/graph-api/} \label{facebook-api}
    \item \url{https://web.archive.org/web/20231123193950/https://www.reddit.com/dev/api/} \label{reddit-api}
    \item \url{https://web.archive.org/web/20231210011040/https://www.redditinc.com/blog/apifacts} \label{reddit-api2}
    \item \url{https://web.archive.org/web/20231210090528/https://www.reddit.com/r/apolloapp/comments/13ws4w3/had_a_call_with_reddit_to_discuss_pricing_bad/} \label{reddit-api3}
    \item \url{https://web.archive.org/web/20231129021701/https://developers.snap.com/creative-kit} \label{snapchat-api}
    \item \url{https://web.archive.org/web/20231212231535/https://api.stackexchange.com/docs} \label{stackexchange-api}
    \item \url{https://web.archive.org/web/20231204034413/https://steamcommunity.com/dev} \label{steam-api}
    \item \url{https://web.archive.org/web/20231210002644/https://developers.tiktok.com/} \label{tiktok-api}
    \item \url{https://web.archive.org/web/20231130070701/https://developer.twitter.com/en/docs/twitter-api}\label{twitter-api}
    \item \url{https://web.archive.org/web/20240303190109/https://developers.google.com/youtube/v3/docs/search/list} \label{youtube-api}
    \item \url{https://web.archive.org/web/20231128205404/https://developers.tiktok.com/products/research-api/} \label{tiktok-api-lim}
    \item \url{https://web.archive.org/web/20231210080026/https://fort.fb.com/researcher-apis} \label{fb-api-lim}
    \item \url{https://web.archive.org/web/20240512003912/https://www.reddit.com/r/AskAcademia/comments/1b32i9q/accessing_reddit_data_for_academic_purposes/?rdt=37952} \label{reddit-api-lim}
    \item \url{https://www.statista.com/statistics/272014/global-social-networks-ranked-by-number-of-users/} \label{link:statista-mau}
    \item \url{https://en.wikipedia.org/wiki/Wikipedia:Wikipedians} \label{link:wikipedia-mau}
    \item \url{https://photutorial.com/flickr-statistics/} \label{link:flickr-mau}
    \item \url{https://thesmallbusinessblog.net/how-many-people-use-yelp/} \label{link:yelp-mau}
    \item \url{https://business.quora.com/resources/reach-over-400-million-monthly-unique-visitors-on-quora/} 
    \label{link:quora-mau}
    \item \url{https://www.statista.com/forecasts/1309791/reddit-mau-worldwide} \label{link:reddit-mau}
    \item \url{https://perspectiveapi.com/research/}\label{link:perspectiveAPI}
    \item \url{https://web.archive.org/web/*/https://www.alexa.com/topsites} \label{link:alexa-top500}
    \item \url{https://en.wikipedia.org/wiki/List_of_social_networking_services} \label{link:wikipedia-osn-list}
    \item \url{https://en.wikipedia.org/wiki/Alt-tech} \label{link:wikipedia-alttech-osn-list}
    \item \url{https://www.statista.com/statistics/1337525/us-distribution-leading-social-media-platforms-by-age-group/} \label{statistaUSSocial}
    \item \url{https://radar.cloudflare.com/domains} \label{link:cloudflare-ranking}
    \item \url{https://www.go-fair.org/fair-principles/}\label{fair}
    \item \url{https://scholar.google.com/citations?view_op=top_venues&hl=en&vq=eng} \label{google-scholar-cats}
    \item \url{https://www.statista.com/statistics/1314183/youtube-shorts-performance-worldwide/} \label{link:youtube}
    \item \url{https://www.statista.com/statistics/795303/china-mau-of-sina-weibo/} \label{link:weibo}
    \item \url{https://www.statista.com/statistics/234038/telegram-messenger-mau-users/} \label{link:telegram}
    \item \url{https://usesignhouse.com/blog/stack-overflow-stats/#:~:text=How%20many%20registered%20users%20of%20Stack%20Overflow%20are%20there%3F,-Want%20a%20link&text=As%20of%20November%202022%2C%20there%20are%20more%20than%20100%20million,and%2023%20million%20registered%20users.} \label{link:stack}
    \item \url{https://web.archive.org/web/2/https://www.nytimes.com/2023/04/18/technology/reddit-ai-openai-google.html} \label{link:reddit-ai}
\end{enumerate}

\end{document}